\documentclass[preprints,article,accept,pdftex,moreauthors]{Definitions/mdpi} 
\firstpage{1} 
\pubvolume{1}
\issuenum{1}
\articlenumber{0}
\pubyear{2022}
\copyrightyear{2022}
\externaleditor{{Academic Editor: Panos Argyrakis} 
}
\datereceived{21 September 2022} 
\dateaccepted{26 October 2022} 
\datepublished{} 
\hreflink{https://doi.org/} % If needed use \linebreak
\usepackage{url}
\usepackage{booktabs}
\usepackage{graphicx}
\usepackage{longtable}
\usepackage[bottom]{footmisc}
\usepackage{caption}
\usepackage{enumitem}
\usepackage{float}
\usepackage[labelformat=simple]{subcaption}

\usepackage{xcolor}
\Title{Dependency Structures in Cryptocurrency Market from High to Low Frequency}

\TitleCitation{Dependency Structures in Cryptocurrency Market from High to Low Frequency}

\Author{Antonio Briola %MDPI: Please carefully check the accuracy of names and affiliations. CORRECT.
 $^{1,2}$ and Tomaso Aste $^{1,2,3,}$*}

\AuthorNames{Antonio Briola, Tomaso Aste}

\AuthorCitation{Briola, A.; Aste, T.}
\address{%
$^{1}$ \quad Department of Computer Science,  University College London,  {London WC1E 6BT}, %MDPI: please confirm the postcode,same below. CORRECT. WE REMOVED THE ANTONIO BRIOLA'S EMAIL SINCE THE AFFILIATION IS THE SAME FOR BOTH AUTHORS AND THE CORRESPONDING AUTHOR IS TOMASO ASTE
 UK\\
$^{2}$ \quad Center for Blockchain Technologies, University College London, {London WC1E 6BT}, UK\\
$^{3}$ \quad Systemic Risk Center, London School of Economics, {London WC2A 2AE}, UK}

\corres{Correspondence: t.aste@ucl.ac.uk}

\abstract{We investigate logarithmic price returns cross-correlations at different time horizons for a set of 25 liquid cryptocurrencies traded on the FTX digital currency exchange. We study how the structure of the Minimum Spanning Tree (MST) and the Triangulated Maximally Filtered Graph (TMFG) evolve from high (15 s) to low (1 day) frequency time resolutions. For each horizon, we test the stability, statistical significance and economic meaningfulness of the networks. Results give a deep insight into the evolutionary process of the time dependent hierarchical organization of the system under analysis. A decrease in correlation between pairs of cryptocurrencies is observed for finer time sampling resolutions. A growing structure emerges for coarser ones, highlighting multiple changes in the hierarchical reference role played by mainstream cryptocurrencies. This effect is studied both in its pairwise realizations and intra-sector ones.}


\begin{document}

\section{Introduction} \label{sec:Introduction}
Financial markets are complex systems~\cite{anderson2018economy}. The~main source of complexity comes from the intricate interaction of heterogeneous actors following various strategies designed to impact at different time scales. They are highly stochastic environments with a low signal to noise ratio, dominated by strong non-stationary dynamics and characterized by feedback loops and non-linear effects~\cite{comerton2015dark, briola2021deep, briola2020deep}. Despite their complexity, financial systems are governed by a rather stable and partially identified framework of rules~\cite{lux1999scaling}. This last characteristic, jointly with the possibility to continuously monitor them across time, makes financial systems well suited for statistical characterization~\cite{aste2010correlation} and a good playground for the study of complex systems in general. In~this paper, we analyse the behaviour of cryptocurrency market. A~cryptocurrency is defined as a digital instrument for value transfer that exploits cryptography and distributed ledgers for security and decentralization~\cite{goodell2022tokens}. As~currencies, they have properties similar to fiat currencies~\cite{fang2022cryptocurrency}. The~main differences being the exclusion of financial institutions as intermediaries~\cite{harwick2016cryptocurrency} and not being controlled and regulated by any central authority~\cite{rose2015evolution}. Thanks to the above mentioned characteristics, cryptocurrency market is available 24 h a day, 7 days a week, and transactions take place between individuals with different physical locations across the globe~\cite{fang2022cryptocurrency}. Standard features of financial systems joined with peculiarities listed above, make cryptocurrencies highly volatile instruments. Finding assets with similar behaviours responding to endogenous or exogenous events is, hence, a challenging, but~extremely valuable, exercise both from theoretical and applicative perspectives (e.g., risk management and investment). The~ready access availability of large volumes of market data ease research on these instruments with respect to classical financial ones. Indeed, one of the main limits faced by research in the field of financial applications is the lack of easy access and share of high-quality data. In~most cases they are sensible data, owned and managed by private financial institutions. Cryptocurrencies are traded on digital currency exchanges (DCEs) which, differently from traditional exchanges, allow to easily access both online and historical data. Exploiting this and using instruments provided by network science, one can successfully build models able to capture and describe individual and collective behaviours in cryptocurrency~market.

A network (or graph) represents components of a system as nodes (or vertices) and interactions among them as links (or edges). The~number of nodes defines the size of the network. The~number of links defines the sparsity (or, conversely, density) of the network. Reversible interactions between components are represented through undirected links, while non-reversible interactions are represented as directed links. Networks have been successfully used in many application domains. Some examples are social networks~\cite{wasserman1994social, newman2002random}, security~\cite{ronfeldt2001networks}, epidemiology~\cite{balcan2009seasonal, hufnagel2004forecast}, neuroscience~\cite{sporns2005human}, drug design~\cite{hopkins2007network}, management~\cite{wu2008mining}, and economic forecasting and modelling~\cite{mantegna1999hierarchical, bonanno2003topology, bonanno2004networks, aste2010correlation, bonanno2001high, wang2021dynamic, procacci2021portfolio, wang2022sparsification}. Many of the above cited works share the peculiarity to study networks with a size varying as a function of time. In~the current research work, on~the contrary, we focus on networks with a fixed size. Starting from a set of 25 liquid cryptocurrencies, we exploit the power of state-of-the-art network-based information filtering approaches (i.e., MST~\cite{west2001introduction} and TMFG~\cite{massara2017network}) to build robust models capturing strong interactions among assets and pruning, at~the same time, weakest ones. We hence investigate dependency structures of the networks at 6 different time horizons spanning from 15 s to 1 day. For~each time horizon, we test the stability, statistical significance, and economic meaningfulness of the graphs. Such a research effort is led by two main motivations. The~first one is the will to describe core dependency structures of the cryptocurrency market in a systematic way, providing a detailed characterization of the reference role played both by mainstream cryptocurrencies and by peripheral ones. The~second is related to the possibility to do this at a wide range of time scales including intra-minute resolutions. Such a characterization is relevant for many reasons. Cryptocurrency market is affected by daily changes related to the introduction of new coins, collapse of existing ones, updates on existing protocols, etc.\ Having a stable framework able to robustly handle this intrinsic mutability, highly eases investment, and risk management decisions and provide a ductile instrument for research purposes. Such a framework should be also able to handle dynamics of cryptocurrencies showing similar characteristics and behaviours (i.e., belonging to the same sector). Dependency structures are, hence, investigated, both at an intra-sector and pairwise level. Unfortunately, there is no consensus on a unique mapping between cryptocurrencies and sectors. We adopt the taxonomy proposed by Kraken~\cite{Kraken} digital currency exchange. Results give a deep insight into the evolutionary process of the time dependent hierarchical organization of the chosen system of cryptocurrencies. As~a further step toward robustness, we compare our results with the ones achieved in the past 20 years of similar research in the field of stock market, uncovering comparable behaviours between the two systems. {From an economic and financial perspective, the~study of dependency structures among cryptocurrencies at different time-scales is relevant both from a theoretical and an applicative point of view. In~the first case, comparing properties of time dependent hierarchical organization of the cryptocurrency market (a relatively young market) with the ones of the equity market (a consolidated market), (i) allows to measure its degree of maturity (ii) keeping track, at~the same time, of~the main evolutionary phases. In~the second case, such an analysis is useful as a support instrument toward the achievement of different goals spanning from portfolio construction tasks (e.g., diversification purposes) to development of multi-assets trading strategies acting at different time-scales.} Our contribution to the existing literature is threefold: (i) we are the first to use TMFG as an information filtering approach to model dependency structures among crypto-assets, (ii) we propose a rigorous network-based study of cryptocurrency market allowing to compare emerging dynamics to the ones observed on traditional financial markets (e.g., the ``Epps effect''), and (iii)~we are the first to describe the evolution of dependency structures among cryptocurrencies at time scales spanning from intra-minute to daily~resolution.

The rest of the paper is organised as follows. In~Section~\ref{sec:Related_Work}, we review the previous research on applications of network science to financial systems modelling. In~Section~\ref{sec:Data}, we discuss the data acquisition and transformation pipeline. In~Section~\ref{sec:Minimum_spanning_tree}, ~Section~\ref{sec:Planar_maximally_filtered_graph}, and in Section~\ref{sec:Triangulated_maximally_filtered_graph}, we characterise cross-correlation between cryptocurrencies as a measure of similarity and dependency. We show how to obtain a dissimilarity measure based on cross-correlation and we review the building process and properties of MSTs, PMFGs and TMFGs. In~Section~\ref{sec:Results}, \textls[-25]{we present results obtained applying methods reported in Section~\ref{sec:Methods}. In~Section~\ref{sec:Conclusions}, we conclude by discussing the economic and financial interpretation of our~findings}.

\section{Related~Work} \label{sec:Related_Work}
Networks have been extensively used in order to model economic and financial systems. The~work by~\cite{mantegna1999hierarchical} can be identified as a foundational one. It demonstrates the possibility to find a hierarchical arrangement of stocks traded in a financial market by investigating the daily time series of logarithmic price returns. A~graph is obtained, exploiting information contained in the correlation matrix computed between all pairs of stocks of the portfolio by considering the synchronous time evolution of the logarithmic returns. Building on the work by~\cite{mantegna1999hierarchical}, the~paper by~\cite{bonanno2000taxonomy} shows that sets of stock index time series can be used to extract meaningful information about the links between different economies across the world. This goal is successfully achieved provided that the effects of the non-synchronous nature of the time series and of the different currencies used to compute the indices are properly taken into account. The~work by~\cite{bonanno2001high} further extends the research by~\cite{mantegna1999hierarchical}, studying modifications of the hierarchical organization of a set of stocks switching from high- to low-frequency time scales. As~a first step, authors report a decrease in correlation between pairs of assets switching from coarser to finer time sampling resolutions. Such a phenomenon is known as ``Epps effect''~\cite{epps1979comovements}. This analysis is extended, investigating both pairwise and intra-sector dynamics. They show the emergence of a more complex network structure at coarser time sampling resolutions, highlighting multiple changes in the hierarchical reference role played by sectors' representative assets. The~work by~\cite{bonanno2003topology} tests the robustness of the findings of the previously cited research works for longer periods of investigation and  demonstrates that networks describing the financial domain cannot be reproduced by a random market model~\cite{laloux1999noise, plerou2002random} and by the one-factor model~\cite{campbell1997econometrics}. Such results are also investigated in~\cite{bonanno2004networks} which specifically shows how the topology of the networks in financial systems can be used to validate or falsify simple, although~widespread, market models. This work also extends the previously cited ones introducing an analysis of the networks built on the volatility of financial time series. More recently, the~work by~\cite{aste2010correlation} shows vulnerabilities of MST~\cite{west2001introduction}  in representing complex systems and proposes the usage of a planar graph, the~PMFG~\cite{aste2005complex}, as~an alternative. This research work also presents a set of methods to validate the statistical significance and robustness of achieved empirical results. The~centrality role of specific financial sectors is finally investigated and the evolution of the Financial sector as a reference one is analysed over a period of 10 years. Recently, some of the network-based information filtering approaches have been sparsely applied to the cryptocurrency market. Results consistent with the ones described in our paper have been recently described by~\cite{vidal2021entry}, adopting a different methodology. In~this research, exploiting the index cohesive force~\cite{kenett2011index}, the~author describes the changes in the hierarchical order of the most influential cryptocurrencies over a period of five years. He shows how Ethereum gradually becomes the most influential cryptocurrency at the detriment of Bitcoin. It is also useful to mention the work by~\cite{zikeba2019shock}, where, for~the first time, the~authors suggest a network-based approach to study the interdependencies between log-returns of cryptocurrencies, with~a special focus on  Bitcoin. They use the MST method in order to group assets into hierarchical clusters and they highlight the potential existence of topological properties of the cryptocurrency market. This work is extended by~\cite{katsiampa2021high}, where, the~authors adopt the MST and the PMFG to study the change in cryptocurrency market's network structure before and after the COVID-19 outbreak. The~last work to be mentioned is the one by~\cite{vidal2022all}, where the author points out how most of the studies on cryptocurrency market are focused only on daily data without considering other options. Using a range of frequencies spanning from one minute to weekly data, he shows how it is possible to detect different profitable frequencies and underlines the relevance of analysing frequencies different from daily~ones. 

\section{Methods} \label{sec:Methods}

\subsection{Data}  \label{sec:Data}
The vast majority of digital currency exchanges provide a free Rest API (or a Web-socket) allowing users to access both historical OHLCV (open, high, low, close, volume) data and online Limit Order Book- and trades-related data. In~addition to this, there is a growing number of services providing out of the box, unified APIs which support many exchanges and merchant APIs. The~work by~\cite{fang2022cryptocurrency} reports a comprehensive and detailed overview of the services currently available for data retrieving. In~the current work, we use data from the FTX~\cite{FTX} digital currency exchange. They are entirely accessed through the CCXT~\cite{CCXT} Python package. We use OHLCV data for 25 cryptocurrencies (see Table~\ref{tab:Crypto_Table}) sampled at time horizons $\Delta t \in [$15 s, 1 min, 15 min, 1 h, 4 h, 1 day$]$. {For each time horizon, a~sample can be defined as a ``time bin''. Opening and closing prices are, respectively, the~first and the last price of the time bin, high and low price are, respectively, the~highest and the lowest price of the time bin and can technically happen in any order, and the~volume is is defined as the sum of the volumes traded in the time bin.} In the rest of the paper, we will use a second-based definition of time horizons. This means that we will refer them as $\Delta t \in [$15, 60, 900, 3600, \text{14,400}, \text{86,400}$]$. Qualitatively, we will often speak about finer and coarser time horizons. In~the first case, we want to indicate elements nearer to the lower bound of the set of time sampling resolutions, while, in~second case, we want to indicate elements nearer to the upper bound of the set of time sampling~resolutions.
\begin{table}[H] 
\tablesize{\small}
\caption{List of the 25 cryptocurrencies analysed in this paper. For~each asset, the~name, the~symbol, the~market capitalization at 29 March 2022 and the corresponding sector according to the taxonomy proposed by~\cite{Messari} is reported. There is no consensus on a unique mapping between cryptocurrencies and sectors. The~chosen taxonomy is the one adopted by one of the main DCEs: Kraken~\cite{Kraken}. Looking at the market capitalization column, it is worth noting that the least capitalized asset is Cream ($\$31.68$M), while the most capitalized one is Bitcoin ($\$903$B). Sectors' grouping is balanced. Cryptocurrencies being the only representative of a specific sector are grouped together in analyses reported in Appendices \ref{Appendix_A} and \ref{Appendix_B}.\label{tab:Crypto_Table}}
\setlength{\cellWidtha}{\textwidth/4-2\tabcolsep+0.2in}
\setlength{\cellWidthb}{\textwidth/4-2\tabcolsep-0.5in}
\setlength{\cellWidthc}{\textwidth/4-2\tabcolsep-0.3in}
\setlength{\cellWidthd}{\textwidth/4-2\tabcolsep+0.6in}
\scalebox{1}[1]{\begin{tabularx}{\textwidth}{>{\centering\arraybackslash}m{\cellWidtha}>{\centering\arraybackslash}m{\cellWidthb}>{\centering\arraybackslash}m{\cellWidthc}>{\centering\arraybackslash}m{\cellWidthd}}

\toprule
\textbf{Cryptocurrency} & \textbf{Symbol} & \textbf{Capitalization} & \textbf{Sector} \\ \midrule
	
	Aave & AAVE & \$2.47B & Lending \\
	Bitcoin Cash & BCH & \$7.13B & Currencies \\
	Binance Coin & BNB & \$72.17B & Centralized~Exchanges \\
	Bitcoin & BTC & \$903B & Currencies \\
	Cream & CREAM & \$31.68M & Lending \\
	Ethereum & ETH & \$412B & Smart Contract~Platforms \\
	FTX Token & FTT & \$7.11B & Centralized~Exchanges \\
	Helium & HNT & \$2.78B & IoT \\
	Huobi Token & HT & \$1.46B & Centralized~Exchanges \\
	Hxro & HXRO & \$129M & Centralized~Exchanges \\
	Litecoin & LTC & \$9.11B & Currencies \\
	Polygon & MATIC & \$13.21B & Scaling \\
	Maker & MKR & \$2.10B & Lending \\
	OMG Network & OMG & \$818M & Scaling \\
	PAX Gold & PAXG & \$609M & Stablecoins \\
	THORChain & RUNE & \$3.96B & Decentralized~Exchanges \\
	Solana & SOL & \$36.09B & Smart Contract~Platforms \\
	Serum & SRM & \$458M & Decentralized~Exchanges \\
	SushiSwap & SUSHI & \$521M & Decentralized~Exchanges \\
	Swipe & SXP & \$800M & Payment~Platforms \\
	TRON & TRX & \$7.24B & Smart Contract~Platforms \\
	 \bottomrule
\end{tabularx}}  \end{table}

\begin{table}[H]\ContinuedFloat
\tablesize{\small}
\caption{{\em Cont.}}
\setlength{\cellWidtha}{\textwidth/4-2\tabcolsep+0.2in}
\setlength{\cellWidthb}{\textwidth/4-2\tabcolsep-0.5in}
\setlength{\cellWidthc}{\textwidth/4-2\tabcolsep-0.3in}
\setlength{\cellWidthd}{\textwidth/4-2\tabcolsep+0.6in}
\scalebox{1}[1]{\begin{tabularx}{\textwidth}{>{\centering\arraybackslash}m{\cellWidtha}>{\centering\arraybackslash}m{\cellWidthb}>{\centering\arraybackslash}m{\cellWidthc}>{\centering\arraybackslash}m{\cellWidthd}}

\toprule
\textbf{Cryptocurrency} & \textbf{Symbol} & \textbf{Capitalization} & \textbf{Sector} \\ \midrule

	Tether & USDT & \$81.37B & Currencies \\
	Waves & WAVES & \$5.77B & Smart Contract~Platforms \\
	XRP & XRP & \$42.05B & Currencies \\
	yearn.finance & YFI & \$836M & Asset Management \\
\bottomrule
\end{tabularx}}
\end{table}

All the considered cryptocurrencies are liquid with a medium-to-high market capitalization. An~exception is Cream, which has a low capitalisation. The~only constraint in the selection process of cryptocurrencies is their historical availability on the FTX digital currency exchange. Indeed, it is worth noting that each digital currency exchange allows to access historical data only starting from the date a specific asset has been quoted on the exchange itself. The~period under analysis spans between 1 January  2021 to 28 February  2022. Despite the high-quality of data, rare missing values are detected at the finest time sampling resolution (i.e., $\Delta t = 15$). In~this case, they are filled using the nearest valid observation. Logarithmic returns (named in the rest of the paper as log-returns) $x$ of closing prices $p$ at time $t$ for a given cryptocurrency $c$, are computed as follows:
\begin{equation} \label{eqn:Log_Rets}
	x_c(t) = \log(p_c(t)) - \log(p_c(t - \Delta t)).
\end{equation}

The assumption of returns' stationarity is validated for each $x_c(t)$ through the Augmented Dickey Fuller (ADF) \cite{dickey1979distribution} test.

\subsection{Correlation-Based~Filtering} \label{sec:Correlation_based_filtering}
Understanding how variables evolve, influencing the collective behaviour, and how the resulting system influences single variables is one of the most challenging problems in complex systems. In~order to extract such an information from the set of synchronous time series discussed in Section~\ref{sec:Data}, we proceed by determining their Pearson's correlation coefficient at each time horizon $\Delta t$. The Pearson's %please check intended meaning is retained. CORRECT.
estimator of the correlation coefficient, for~non-overlapping increments, between~two synchronous data series with length $T\Delta t$ is:
\begin{equation}
	{\rho_{i,j}(\Delta t) = \frac{\frac{1}{T} \sum_{u=1}^T (x_i(u \Delta t) - \mu_i)(x_j(u \Delta t) - \mu_j)}{\sigma_i \sigma_j}}
\end{equation}
where $\mu_{i(j)}$ and $\sigma_{i(j)}$ are, respectively, the~sample mean and the sample standard deviation of the data series $x_{i(j)}(t)$. The~Pearson’s correlation coefficient is a widespread measure efficient at catching similarities between the evolution process of financial assets' prices~\cite{aste2010correlation}. By~definition, $\rho_{i,j}(\Delta t)$ has values between $-1$ (meaning that the two synchronous time series are completely, linearly anti-correlated) and $+1$ (meaning that the two synchronous time series are completely, linearly correlated). When $\rho_{i,j}(\Delta t) = 0$, the~two synchronous time series are linearly uncorrelated. The~correlation matrix $\textbf{C}$ %MDPI: Is the bold necessary? please check it through the whole maintext. IT IS NECESSARY AND MAINTAINED THROUGH THE WHOLE TEXT.
 is $n \times n$ (where \textit{n} is the number of variables) symmetric, with~elements on the diagonal equal to one (i.e., $\rho_{i,i}(\Delta t) = 1$). For~each time horizon $\Delta t$, $n(n-1)/2$ correlation coefficients completely characterize the correlation matrix. From~a network science perspective, the~correlation matrix can be considered as a fully connected graph where each asset is represented by a node and each pair of assets is joined by an undirected edge representing their~correlation.

\subsection{Minimum Spanning Tree (MST)} \label{sec:Minimum_spanning_tree}
Based on the correlation matrix, we want to build an undirected graph whose topology captures dependency structures among cryptocurrencies' log-returns time series and that is greatly reduced in the number of edges with respect to a complete graph. In~such a network, all the relevant relations must be represented. At~the same time, the~network should be kept as simple as possible. The~simplest connected graph is a spanning tree. Minimum spanning trees (MSTs) \cite{west2001introduction} are largely used in multivariate analysis; they represent a class of networks that connect all the vertices without forming cycles (i.e., closed paths of at least three nodes). MSTs are often computed with respect to a distance metric, so that minimizing the metric corresponds to linking assets that are close to each other. As~a product of their building process, MSTs retain the maximum possible number of distances~\cite{mantegna1999hierarchical} minimizing, at~the same time, the~total edge distance. In~\cite{mantegna1999hierarchical}, MSTs are computed using the Euclidean distance~\cite{gower1966some}:
\begin{equation}\label{eq:Gower_Distance}
    d_{i,j} = \sqrt{2(1-\rho_{i,j})}.
\end{equation}

This definition is however too restrictive disfavouring negatively correlated variables that are equally important as the positive ones for the representation of the dependency structure~\cite{soramaki2016network}. In~order to mitigate this limitation, we use the power dissimilarity measure:
\begin{equation}\label{eq:Square_Distance}
    d_{i,j} = 1-\rho_{i,j}^2
\end{equation}

The work~\cite{mantegna1999introduction} provides a complete pedagogical exposition of the determination of the MST in the context of synchronous financial time series. A~general approach to the construction of the MST is to connect the less dissimilar vertices while constraining the graph to be a tree as follows:

\begin{enumerate}
	\item Make an ordered list of edges $i, j$, ranking them by increasing dissimilarity (first the edge expressing the highest similarity and last the edge expressing the highest dissimilarity).
	\item Pop the first element of the ordered list and add it to the spanning tree.
	\item If the added edge creates a cycle then remove the edge, otherwise skip to step \ref{item:MST_4}.
	\item Iterate the process from step 2 until all pairs have been exhausted. \label{item:MST_4}
\end{enumerate}

Such an algorithm for the construction of the MST is known as the Prim's algorithm~\cite{prim1957shortest}. The~resulting network has $n-1$ edges. Considering that the system of cryptocurrencies analysed in the current paper is made of $n = 25$ assets (i.e., nodes), the~resulting MST contains $24$ edges (the code used to compute MSTs can be retrieved at \url{https://github.com/shazzzm/topcorr}; last access on 27 October 2022.).%MDPI: footnote is not allowed for this journal, we moved it to the maintext, please confirm and add the accessed date for the link. MOVING ACCEPTED AND ACCESS DATE PROVIDED.

\subsection{Planar Maximally Filtered Graph (PMFG)} \label{sec:Planar_maximally_filtered_graph}
The MST is a powerful method to capture meaningful relationships in a network structure describing a complex system. However, this method presents some aspects that can be unsatisfactory. The~main constraint is that it has to be a tree (i.e., it cannot contain cycles). This characteristic makes impossible to represent relationships among more than two variables showing strongly correlated behaviours in their dynamics. In~order to maintain the same powerful filtering properties of the MST and adding, at~the same time, extra links, cycles, and cliques (i.e., complete subgraphs) in a controlled manner, it has been proposed to use the Planar Maximally Filtered Graph (PMFG) \cite{tumminello2005tool, aste2006dynamical, tumminello2007correlation, di2010use}. PMFG can be viewed as the first incremental step towards complexity after the MST. Indeed, instead of being a tree, the~algorithm impose planarity. A~graph is said to be planar if it can be embedded in a sphere without edges crossing. The~foundational work by~\cite{aste2010correlation} provides a comprehensive pedagogical exposition of the determination of the PMFG. A~general approach to the construction of the PMFG can be resumed as follows:

\begin{enumerate}
	\item Make an ordered list of edges $i, j$, ranking them by increasing dissimilarity (first the edge expressing the highest similarity and last the edge expressing the highest dissimilarity).
	\item Pop the first element of the ordered list and add it to the graph.
	\item If the  resulting graph is not planar, then remove the edge, otherwise skip to step \ref{item:PMFG_4}.
	\item Iterate the process from step 2 until all pairs have been exhausted. \label{item:PMFG_4}
\end{enumerate}

It has been proved that the MST is always a sub-graph of the PMFG~\cite{tumminello2005tool}. PMFG has $3 \times (n-2)$ edges and a number of 3-cliques larger or equal to $2n-4$. We remark that also 4-cliques can be present in this kind of~graph.

\subsection{Triangulated Maximally Filtered Graph (TMFG)}
\label{sec:Triangulated_maximally_filtered_graph}
The PMFG presents two main limits: it is computational costly and it is a non-chordal graph. A~graph is said to be chordal if all cycles made of four or more vertices have a chord which reduces the cycle to a set of triangles. A~chord is defined as an edge that is not part of the cycle but connects two vertices of the cycle itself. In~order to bypass these two constraints, the~Triangulated Maximally Filtered Graph (TMFG) \cite{massara2017network} has been proposed. A~general approach to the construction of the TMFG can be resumed as follows:

\begin{enumerate}
\item Make an ordered list of edges $i, j$, ranking them by increasing dissimilarity (first the edge expressing the highest similarity and last the edge expressing the highest dissimilarity).
\item Find the 4 nodes with the lowest sum of edge weights with all other nodes in the graph and connect them forming a tetrahedron with 4 triangular faces.
\item Identify and add the node that minimize the sum of its connections to a triangle face already included in the graph, forming three new triangular faces. \label{item:TMFG_3}
\item If the graph reaches a number of edges equal to $3n - 6$, then stop, otherwise go to \mbox{step \ref{item:TMFG_3}}.
\end{enumerate}

Such an algorithm extracts a planar subgraph which optimises an objective function quantifying the gain of adding a new vertex to the existing tetrahedron. Compared to the PMFG, the~TMFG is more efficient to be computed and is a chordal graph. The~chordal structural form allows to use the filtered graph for probabilistic modeling~\cite{turiel2020simplicial, barfuss2016parsimonious}. A~TMFG has $3 \times (n-2)$ edges (with $n$ representing the number of nodes) and contains both 3-cliques and 4-cliques. Considering that the system of cryptocurrencies analysed in the current paper is made of $n = 25$ assets (i.e., nodes), the~resulting TMFG contains $69$~edges, $88$ $3$-cliques, and $22$ $4$-cliques (The code used to compute TMFGs can be retrieved at \url{https://github.com/shazzzm/topcorr};  last access on 27 October 2022.).

\section{Results}\label{sec:Results}
\begin{figure}[H] %FIGURE ENLARGED
	\hspace{-30pt}\begin{subfigure}{.45\textwidth}
		\centering
		\includegraphics[scale=0.25]{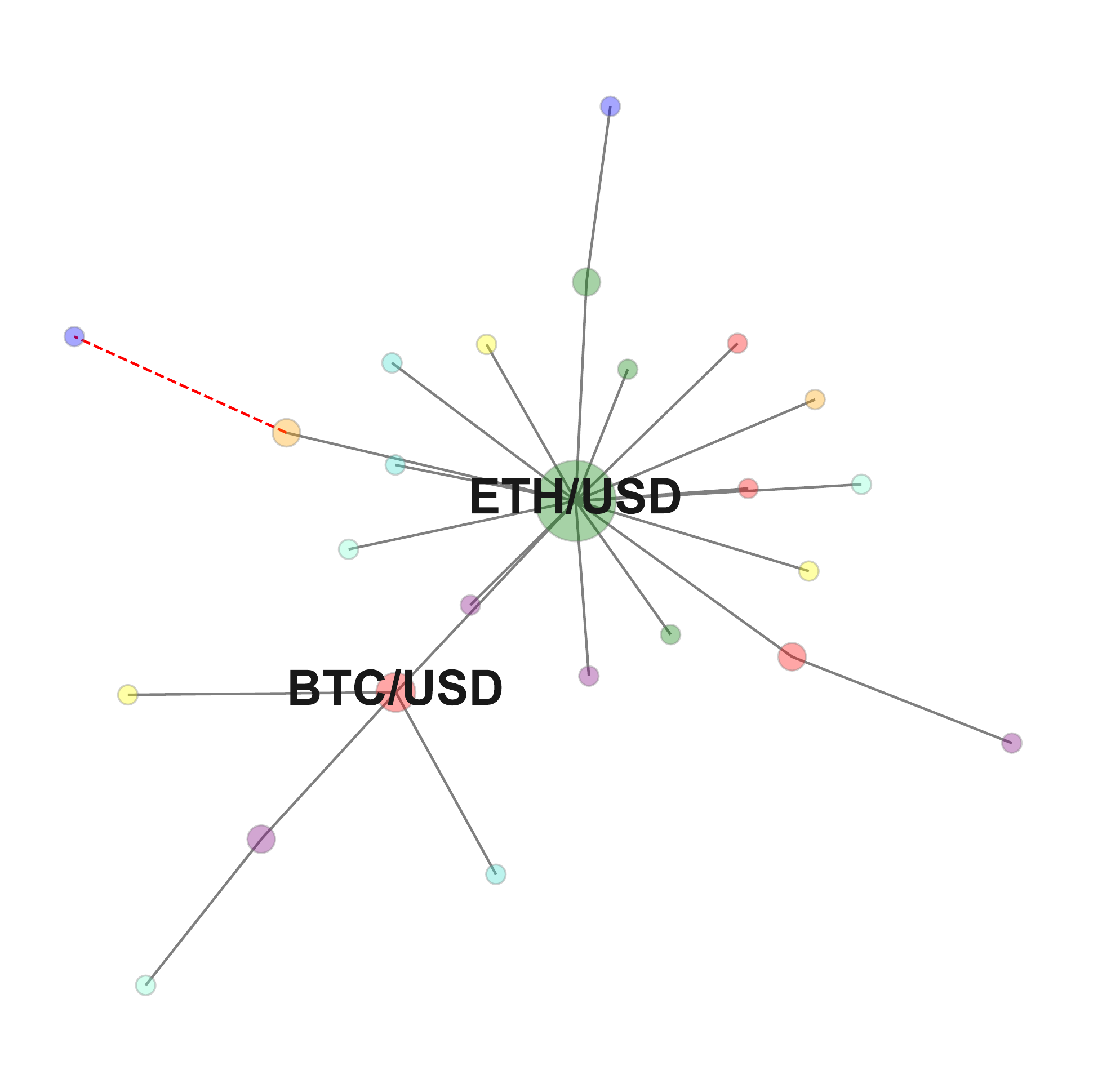}  
		\caption{\centering}
		\label{fig:Intra_Text_MST_coloured_15}
	\end{subfigure}
	\begin{subfigure}{.45\textwidth}
		\centering
		\includegraphics[scale=0.25]{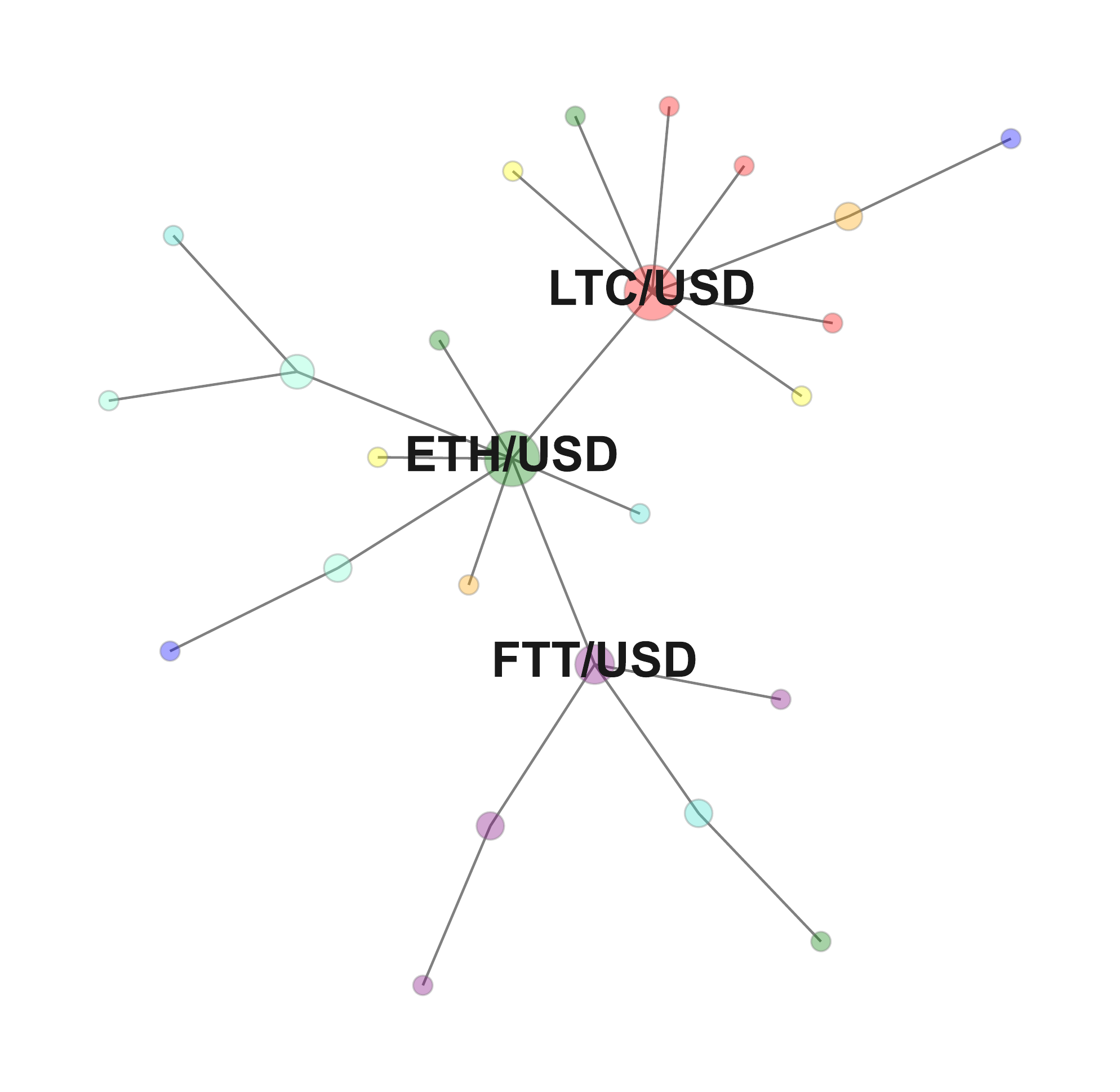}  
		\caption{\centering}
		\label{fig:Intra_Text_MST_coloured_86400}
	\end{subfigure}
	\vspace{2mm}
	\caption{Minimum Spanning Tree representing log-returns time series' dependency structure computed at (\textbf{a}) 15 s and (\textbf{b}) 1 day. Only {hub} nodes are labelled. The~adopted colour mapping scheme follows the sectors' taxonomy by~\cite{Messari} (see Appendix \ref{Appendix_A}).}
	\label{fig:Intra_Text_MST_coloured}
\end{figure}

\begin{figure}[H]
	\hspace{-25pt}\begin{subfigure}{.45\textwidth}
		\centering
		\includegraphics[scale=0.25]{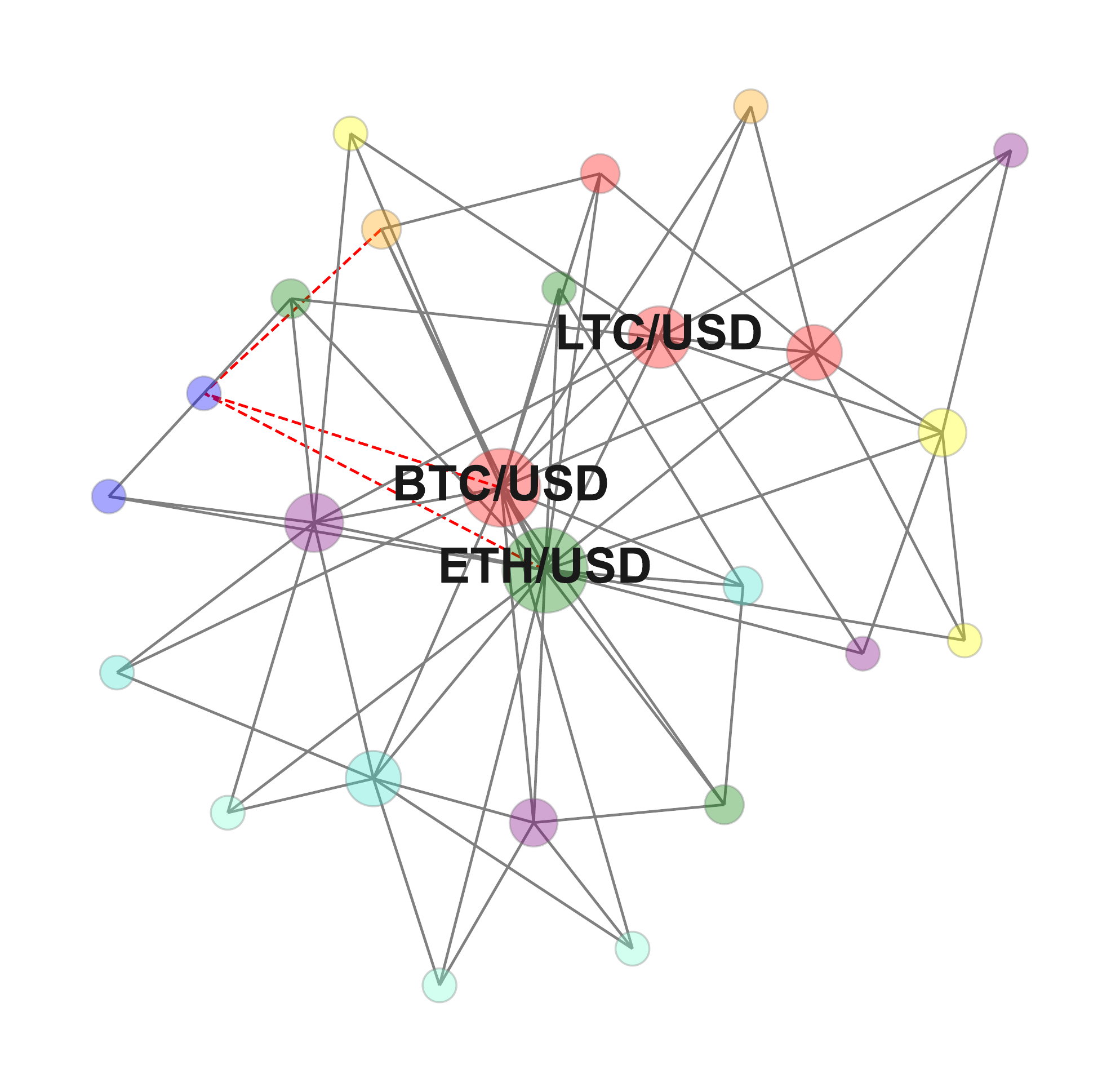}  
		\caption{\centering}
		\label{fig:Intra_Text_TMFG_coloured_15}
	\end{subfigure}
	\begin{subfigure}{.45\textwidth}
		\centering
		\includegraphics[scale=0.25]{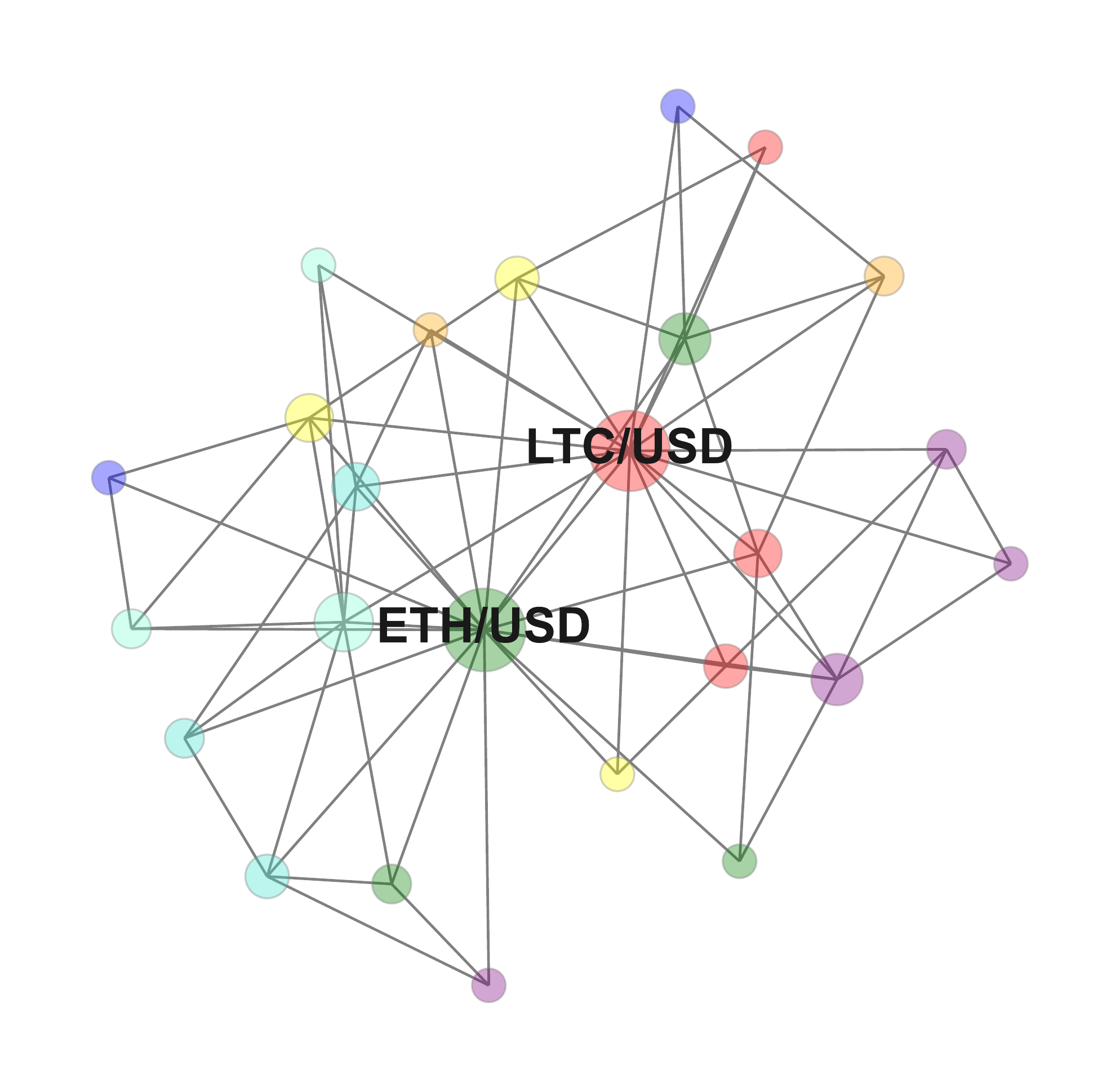}  
		\caption{\centering}
		\label{fig:Intra_Text_TMFG_coloured_86400}
	\end{subfigure}
	\vspace{2mm}
	\caption{Triangulated Maximally Filtered graphs representing log-returns time series' dependency structure computed at (\textbf{a}) 15 s and (\textbf{b}) 1 day. Only {hub} nodes are labelled. The~adopted colour mapping scheme follows the sectors' taxonomy by~\cite{Messari} (see Appendix \ref{Appendix_B}).}
	\label{fig:Intra_Text_TMFG_coloured}
\end{figure}

Figures~\ref{fig:Intra_Text_MST_coloured}a and \ref{fig:Intra_Text_TMFG_coloured}a report the MST and the TMFG computed at horizon $\Delta t = 15$. Figures~\ref{fig:Intra_Text_MST_coloured}b and ~\ref{fig:Intra_Text_TMFG_coloured}b report the MST and the TMFG computed at horizon $\Delta t = \text{86,400}$. Full set of MSTs computed following the procedure described in Section~\ref{sec:Minimum_spanning_tree} is reported in Appendix~\ref{Appendix_A}. Full set of TMFGs computed following the procedure described in Section~\ref{sec:Triangulated_maximally_filtered_graph} is reported in Appendix \ref{Appendix_B}.
\vspace{-10pt}
\unskip

\begin{figure}[H]
	\begin{subfigure}{.45\textwidth}
		\centering
		\includegraphics[width=1\linewidth]{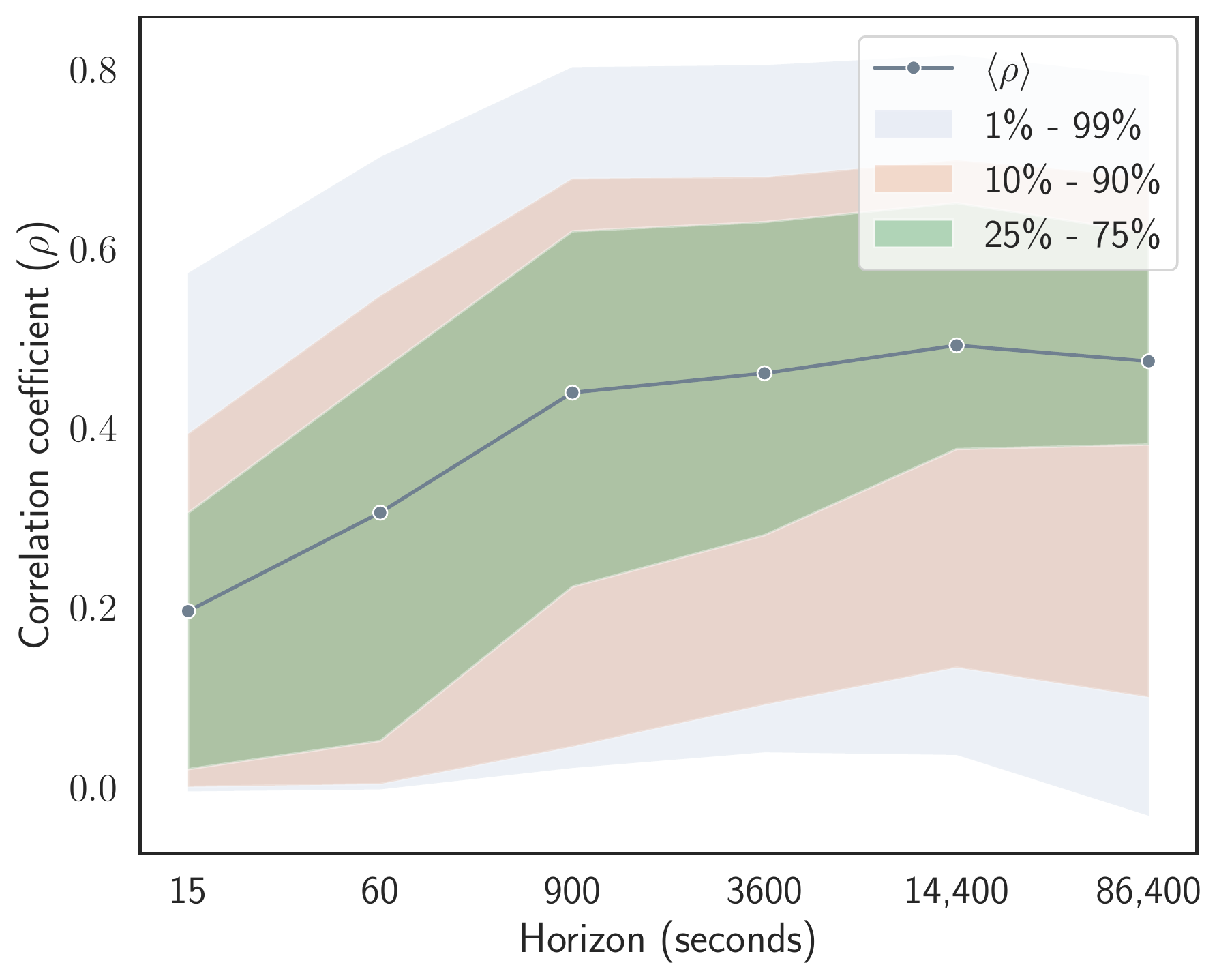}  
		\caption{\centering}
		\label{fig:Inter_Sector_Correlation_Coefficient}
	\end{subfigure}
	\begin{subfigure}{.45\textwidth}
		\centering
		\includegraphics[width=1\linewidth]{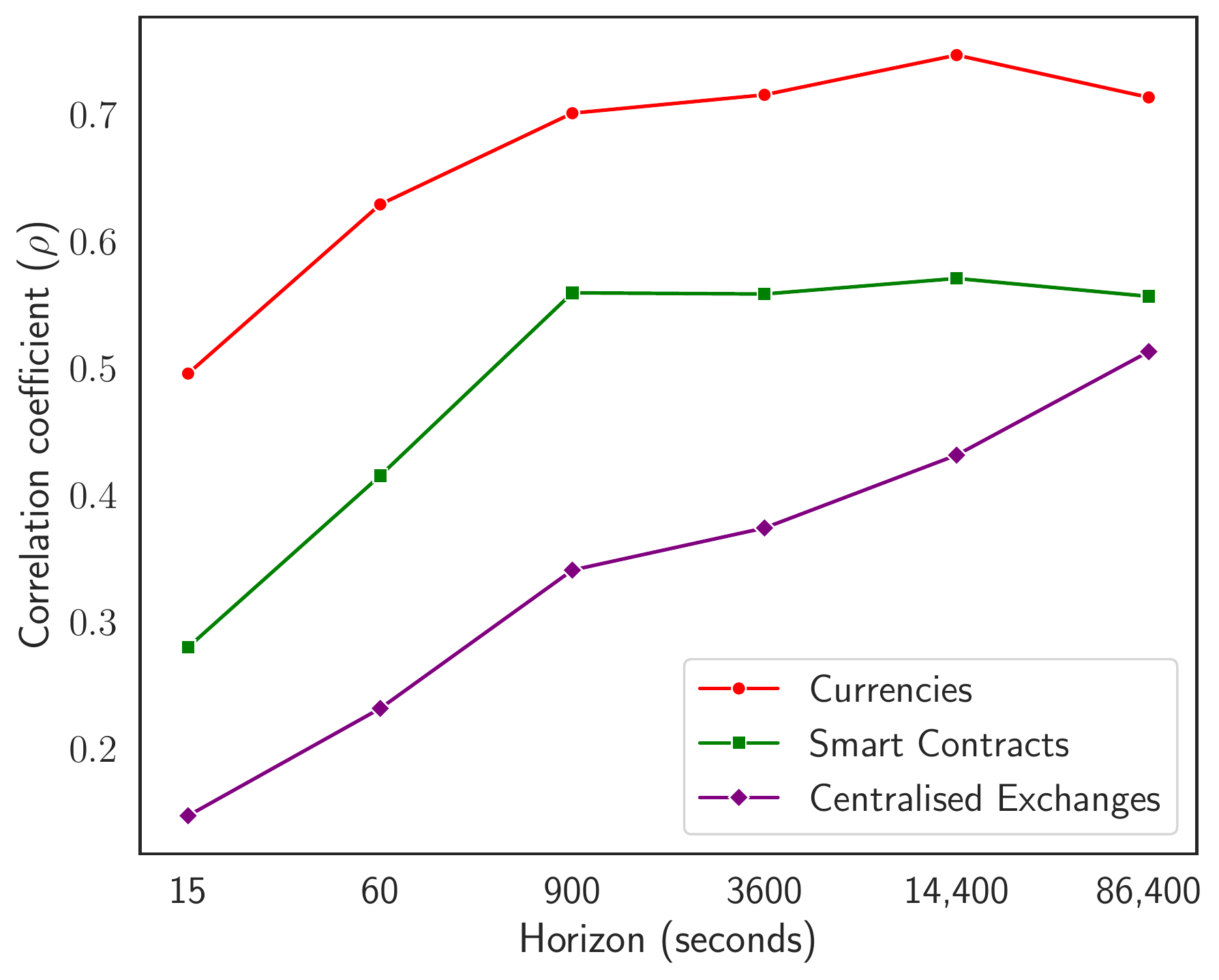}  
		\caption{\centering}
		\label{fig:Intra_Sector_Correlation_Coefficient}
	\end{subfigure}
	\vspace{2mm}
	\caption{Evolutionary dynamics %MDPI: Please use commas to separate thousands for numbers with five or more digits (not four digits) in the picture, e.g., “10000” should be “10,000”. CHANGES DONE
 of the average correlation coefficient as a function of the time horizon $\Delta t$.  (\textbf{a}) reports the horizon-related mean Pearson's correlation coefficient and three meaningful percentiles computed averaging over the $n(n-1)/2 = 300$ off-diagonal elements of the whole correlation matrix \textbf{C}. (\textbf{b}) reports the horizon related mean Pearson's correlation coefficients computed averaging over the $n_s(n_s - 1)/2$ correlation coefficients of the $n_s$ assets belonging to one of three of the most relevant sectors defined by~\cite{Messari}: Currencies, Smart Contracts, Centralised Exchange~sectors.}
	\label{fig:Correlation_Coefficients_Analysis}
\end{figure}

As a preliminary step into the study \textls[-5]{of the information level carried by the two network-based information filtering approaches, Figure~\ref{fig:Correlation_Coefficients_Analysis} shows how pairwise (see \mbox{Figure~\ref{fig:Correlation_Coefficients_Analysis}a})} and the intra-sector (see Figure~\ref{fig:Correlation_Coefficients_Analysis}b) average Pearson's correlation coefficient $\langle \rho \rangle$ evolves as a function of time horizon $\Delta t$. Figure~\ref{fig:Correlation_Coefficients_Analysis}a reports the mean Pearson's correlation coefficient computed averaging over the $n(n-1)/2 = 300$ off-diagonal elements of the whole correlation matrix \textbf{C} at different time horizons. In~order to give a more comprehensive view of the evolutionary dynamics of the mean pairwise correlation coefficient, we also report three meaningful percentile intervals. We observe that the average correlation coefficient $\langle \rho \rangle$ increases with time horizon $\Delta t$ from a value equals to $0.19$ at $\Delta t = 15$ to a value equals to $0.47$ at $\Delta t = \text{86,400}$. The~value at $\Delta t = 15$ corresponds to the minimum average correlation coefficient across time horizons. On~the other hand, the~maximum average correlation coefficient does not coincide with the one computed at the maximum time horizon. It is instead detected at horizon $\Delta t = \text{14,400}$, which corresponds to an intra-day resolution (i.e., 4 h). On~average, the~most prominent pairwise correlation weakenings are observed for most correlated pair of assets (i.e., those pairs of cryptocurrencies having a correlation coefficient included into highest percentiles).

Figure~\ref{fig:Correlation_Coefficients_Analysis}b reports mean Pearson's correlation coefficient computed averaging over the $n_s(n_s - 1)/2$ correlation coefficients of the $n_s$ assets belonging to one specific sector~\cite{Messari} at different time horizons. Specifically, we report dynamics for Currencies, Smart Contracts, and Centralized Exchanges sectors. This choice is completed considering the relevance of the three sectors. The relevance of sectors is defined in relation to results discussed later in this section. An intra-sector scenario shows trends comparable to the ones observed in pairwise context. All the previously discussed dynamics are here more pronounced. In~both cases, we observe the ``Epps effect'', i.e.,~a decrease in pair correlations at finer time sampling resolutions. This effect has been extensively studied in equity markets by~\cite{epps1979comovements, bonanno2001high}. Results reported in Figure~\ref{fig:Correlation_Coefficients_Analysis} show how, also in the cryptocurrency market, the~intra-sector correlation increases faster than pairwise one. The~``Epps effect'' is, hence, more pronounced within each sector than outside it. Going deeper, in~Appendix \ref{Appendix_C}, we compare the probability distribution of correlation coefficients in the empirical correlation matrix \textbf{C} with the probability distribution of correlation coefficients filtered, respectively, by~the MST and by the TMFG at different time horizons. We also report the probability distribution of correlation coefficients for surrogate multivariate time series obtained by randomly shuffling log-returns time series of the $25$ cryptocurrencies listed in Table~\ref{tab:Crypto_Table}. This step is performed in order to evaluate the null hypothesis of uncorrelated returns for the considered portfolio of cryptocurrencies. Results give us the possibility to asses the statistical significance of average correlation coefficients chosen both by MST and by TMFG networks. These findings are reported in a synthetic way in Table~\ref{tab:Average_Correlation_Table}. The~extended count {and the corresponding statistical meaning} of links having a value higher than the minimum and lower than the maximum correlation coefficient detected by shuffling log-returns time series at different time horizons for the three scenarios are reported in Appendix \ref{Appendix_D}.
\begin{table}[H]
\tablesize{\small}
\caption{Average absolute correlation coefficient $\langle |\rho| \rangle$ and quantiles (25--$75\%$) computed on the empirical correlation matrix \textbf{C}, on~the links filtered by MST and on the ones filtered by TMFG at different time horizons. Statistical significance of the average correlation coefficient is represented though asterisks. \emph{p}-values $> 0.05$ are not marked. \emph{p}-values $\leq 0.05$ are marked as $^{\ast}$. \emph{p}-values $\leq 0.01$ are marked as $^{\ast\ast}$. \emph{p}-values $\leq 0.001$ are marked as $^{\ast\ast\ast}$. The~filtering power of the MST and TMFG is evident considering that the related mean correlation coefficients are always greater than the ones computed on the whole correlation coefficient matrix \textbf{C}. Results for both MST and TMFG are always robust across time~horizons.}
\label{tab:Average_Correlation_Table}

\setlength{\cellWidtha}{\textwidth/10-2\tabcolsep-0in}
\setlength{\cellWidthb}{\textwidth/10-2\tabcolsep-0in}
\setlength{\cellWidthc}{\textwidth/10-2\tabcolsep-0in}
\setlength{\cellWidthd}{\textwidth/10-2\tabcolsep-0in}
\setlength{\cellWidthe}{\textwidth/10-2\tabcolsep-0in}
\setlength{\cellWidthf}{\textwidth/10-2\tabcolsep-0in}
\setlength{\cellWidthg}{\textwidth/10-2\tabcolsep-0in}
\setlength{\cellWidthh}{\textwidth/10-2\tabcolsep-0in}
\setlength{\cellWidthi}{\textwidth/10-2\tabcolsep-0in}
\setlength{\cellWidthj}{\textwidth/10-2\tabcolsep-0in}
\scalebox{1}[1]{\begin{tabularx}{\textwidth}{>{\centering\arraybackslash}m{\cellWidtha}>{\centering\arraybackslash}m{\cellWidthb}>{\centering\arraybackslash}m{\cellWidthc}>{\centering\arraybackslash}m{\cellWidthd}>{\centering\arraybackslash}m{\cellWidthe}>{\centering\arraybackslash}m{\cellWidthf}>{\centering\arraybackslash}m{\cellWidthg}>{\centering\arraybackslash}m{\cellWidthh}>{\centering\arraybackslash}m{\cellWidthi}>{\centering\arraybackslash}m{\cellWidthj}}
\toprule

\boldmath{$\Delta t$} & \multicolumn{3}{c}{\textbf{C}}       & \multicolumn{3}{c}{\textbf{MST}}         & \multicolumn{3}{c}{\textbf{TMFG}}         \\ \midrule
 & \textbf{$\langle |\rho| \rangle$} & \textbf{$25\%$} & \textbf{$75\%$} & \textbf{$\langle |\rho| \rangle$} & \textbf{$25\%$} & \textbf{$75\%$} & \textbf{$\langle |\rho| \rangle$} & \textbf{$25\%$} & \textbf{$75\%$} \\ \midrule
15                  & $0.20\;\;\;\;$              & $0.02$ & $0.31$ & $0.35^{\;\ast\ast\ast}$ & $0.26$ & $0.49$ & $0.31^{\;\ast\ast}\;$     & $0.22$ & $0.42$ \\
60                  & $0.31\;\;\;\;$              & $0.05$ & $0.46$ & $0.47^{\;\ast\ast\ast}$ & $0.42$ & $0.63$ & $0.44^{\;\ast\ast\ast}$ & $0.38$ & $0.57$ \\
900                 & $0.44^{\;\ast\ast}$ & $0.22$ & $0.62$ & $0.60^{\;\ast\ast\ast}$ & $0.57$ & $0.76$ & $0.57^{\;\ast\ast\ast}$ & $0.55$ & $0.69$ \\
3600                & $0.46^{\;\ast\ast}$ & $0.28$ & $0.63$ & $0.62^{\;\ast\ast\ast}$ & $0.59$ & $0.76$ & $0.59^{\;\ast\ast\ast}$ & $0.56$ & $0.70$ \\
14,400               & $0.49^{\;\ast}\;$ & $0.38$ & $0.65$ & $0.65^{\;\ast\ast\ast}$ & $0.64$ & $0.77$ & $0.62^{\;\ast\ast\ast}$ & $0.58$ & $0.72$ \\
86,400               & $0.48\;\;\;\;$              & $0.38$ & $0.62$ & $0.66^{\;\ast\ast}\;$     & $0.64$ & $0.77$ & $0.61^{\;\ast\ast}\;$     & $0.57$ & $0.72$ \\

\bottomrule
\end{tabularx}}

\end{table}

Average correlation coefficients for MSTs and TMFGs are always greater than the ones computed on the empirical correlation matrix \textbf{C}. The~difference between cross-horizons mean of average correlation coefficients filtered by MSTs and cross-horizons mean of average correlation coefficients in \textbf{C}, is equal to $0.16$. The~difference between cross-horizons mean of average correlation coefficients filtered by TMFGs and cross-horizons mean of average correlation coefficients in \textbf{C}, is equal to $0.12$. Correlation coefficients filtered by TMFGs are always lower than the ones filtered by MSTs. This depends on the fact that, as~reported in Section~\ref{sec:Triangulated_maximally_filtered_graph}, the~TMFG contains, by~construction, more information than the MST. The~mean difference between average correlation coefficients filtered by MSTs and the ones filtered by TMFGs, is equal to $0.03$. Results reported in Table~\ref{tab:Average_Correlation_Table} confirm that the two filtering approaches prune weakest correlations among considered cryptocurrencies keeping only the strongest ones. Differently from what happens for the empirical correlation matrix \textbf{C}, results for both MST and TMFG are always statistical significant across time horizons. These results enforce the evidence that both MST and TMFG carry information about strongest interactions observed in the system, disregarding most of the links consistent with the null hypothesis of uncorrelated data. It is worth noting that such an analysis does not tell much about the statistical robustness of links selected by the two network-based information filtering approaches. In~order to perform such an investigation, we adopt the technique proposed by~\cite{tumminello2007spanning}. For~each time horizon $\Delta t$, we sample 1000 bootstrap replicas $r = 1, \dots, 1000$ of the empirical log-returns time series data. The~length of empirical data and the one of each replica is kept equal. We compute the MST*(\textit{r}) and the TMFG*(\textit{r}) for each replica \textit{r}. For~each time sampling resolution, we map each link of the original MST and TMFG to an integer number and we count the number of links present both in the MST and TMFG and in each of the  MST*(\textit{r}) and TMFG*(\textit{r}). Table~\ref{tab:Bootstrap_Edges} reports, for~each time sampling interval $\Delta t$, the~number of links of the empirical MST and TMFG with a bootstrap value larger than $95\%$.
\begin{table}[H]
\tablesize{\small}
\caption{Percentage of links contained in empirical MST and TMFG at time horizon $\Delta t$ with a bootstrap value larger than $95\%$. In~the case of the MST, it is possible to notice how the robustness of the network structure decreases for coarser time sampling intervals. In~the case of TMFG, on~the contrary, the~robustness is maintained across time horizons with low~oscillations.}
\label{tab:Bootstrap_Edges}

\setlength{\cellWidtha}{\textwidth/3-2\tabcolsep-0in}
\setlength{\cellWidthb}{\textwidth/3-2\tabcolsep-0in}
\setlength{\cellWidthc}{\textwidth/3-2\tabcolsep-0in}
\scalebox{1}[1]{\begin{tabularx}{\textwidth}{>{\centering\arraybackslash}m{\cellWidtha}>{\centering\arraybackslash}m{\cellWidthb}>{\centering\arraybackslash}m{\cellWidthc}}
\toprule

\boldmath{$\Delta t$} & \textbf{MST} & \textbf{TMFG} \\ \midrule
15                  & $62.5\%$     & $28.9\%$      \\
60                  & $58.3\%$     & $37.7\%$      \\
900                 & $54.2\%$     & $36.2\%$      \\
3600                & $58.3\%$     & $36.2\%$      \\
14,400               & $41.6\%$     & $40.6\%$      \\
86,400               & $25.0\%$     & $27.5\%$      \\

\bottomrule
\end{tabularx}}

\end{table}

%\begin{table}[H]
%\centering
%\caption{Percentage of links contained in empirical MST and TMFG at time horizon $\Delta t$ with a bootstrap value larger than $95\%$. In~the case of the MST, it is possible to notice how the robustness of the network structure decreases for coarser time sampling intervals. In~the case of TMFG, on~the contrary, the~robustness is maintained across time horizons with low~oscillations.}
%\label{tab:Bootstrap_Edges}
%\begin{tabular}{@{}ccc@{}}
%\toprule
%\textbf{$\Delta t$} & \textbf{MST} & \textbf{TMFG} \\ \midrule
%15                  & $62.5\%$     & $28.9\%$      \\
%60                  & $58.3\%$     & $37.7\%$      \\
%900                 & $54.2\%$     & $36.2\%$      \\
%3600                & $58.3\%$     & $36.2\%$      \\
%14400               & $41.6\%$     & $40.6\%$      \\
%86400               & $25.0\%$     & $27.5\%$      \\ \bottomrule
%\end{tabular}
%\end{table}

Results in Table~\ref{tab:Bootstrap_Edges} show how, in~the case of the MST, the~robustness of the underlying network structure decreases for coarser time sampling resolutions. A~consistent result has been observed by~\cite{tumminello2007correlation} in equity markets. This finding can be explained in two different ways. The~first and most straightforward explanation is the statistical one and can be resumed as follows: the higher number of samples at finer time sampling resolutions implies higher statistical significance, while the lower number of samples at coarser time sampling resolutions imply lower statistical significance. A~second explanation can be given looking at the structure of the networks reported in Appendix \ref{Appendix_A}. At~finer time sampling resolutions, we observe less structured networks where numerous small-degree nodes (spokes) coexist with few anchor ones (hubs) characterised by an exceptionally high number of links. At~coarser time sampling resolutions we observe more structured networks with a less imbalanced degree distribution. Such a topological change directly implies a loss in the links' statistical robustness. The~case of TMFG is different. Statistical robustness of the network is maintained across horizons without significant draw-downs. Indeed, during~the optimization phase of the objective function, TMFG tends to be marginally exposed to local minima, being robust to dramatic topological~changes. 

\begin{figure}[H]
	\centering 
	\includegraphics[scale=0.45]{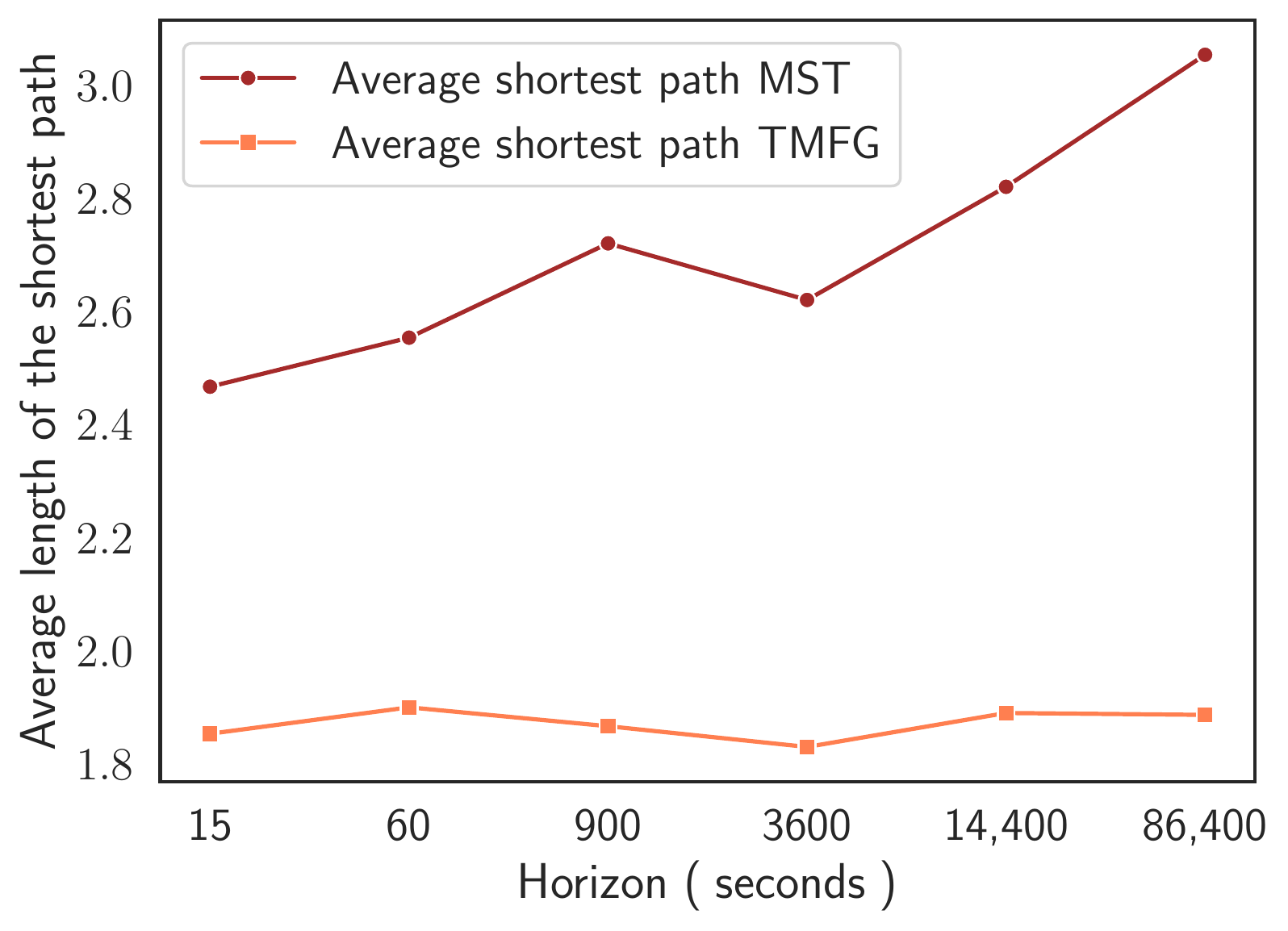}  
	\caption{Average length %MDPI: Please use commas to separate thousands for numbers with five or more digits (not four digits) in the picture, e.g., “10000” should be “10,000”. CHANGE DONE. WE PREFER THE IMAGE TO BE CENTERED.
 of the shortest path in MST and TMFG as function of the time horizon at which log-returns are computed. We observe a decreasing compactness of MST networks at coarser time sampling resolutions. Instead, the~compactness of the TMFG turns out to be stable across time horizons with low~oscillations.}
	\label{fig:shortest_path}
	
\end{figure}

These last findings can be formally characterised studying the evolution of the average shortest path in MST and in TMFG as a function of time sampling resolution. Figure~\ref{fig:shortest_path} reports the significant different behaviour in compactness' evolutionary dynamics of the two network-based information filtering approaches. In~the case of MST, the~minimum length of the average shortest path is equal to $2.46$ and is detected at $\Delta t = 15$, while the maximum length is equal to $3.05$ and is detected at $\Delta t = \text{86,400}$. In~the case of TMFG, we observe a strong compactness across time horizons. The~minimum length of the average shortest path is equal to $1.83$ and is detected at $\Delta t = 3600$, while the maximum length is equal to $1.9$ at $\Delta t = 60$. In~the case of MST, at~the finest time sampling resolution (i.e., $\Delta t = 15$), we observe a structurally simple network with two cryptocurrencies (i.e., Ethereum and Bitcoin) acting as a hierarchical reference for the majority of other assets. This topological structure persists switching to time horizon $\Delta t = 60$. Several changes in nodes' reference roles can be observed for networks sampled at time horizons $\Delta t = 900$ and $\Delta t = 3600$. 

In both cases Ethereum maintains its reference role even reducing its centrality. Bitcoin, on~the contrary, is gradually replaced in its role by Litecoin and FTX token (both part of the Bitcoin's cluster at time horizon $\Delta t = 60$). This structural transition is evident at $\Delta t = \text{14,400}$ and fully realised at $\Delta t = \text{86,400}$.

\begin{figure}[H]
	\begin{subfigure}{.45\textwidth}
		\centering
		\includegraphics[width=1\linewidth]{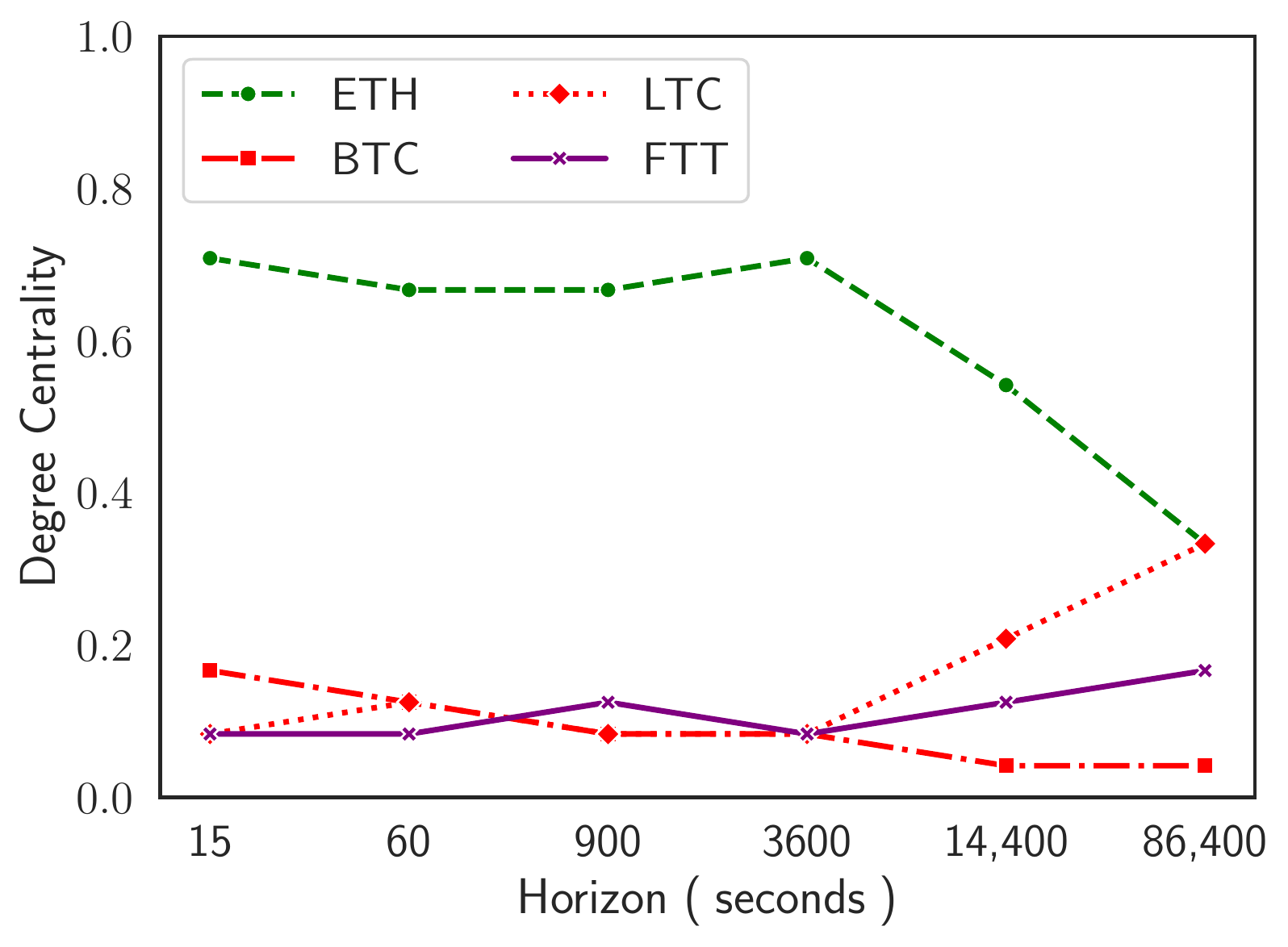}  
		\caption{\centering}
		\label{fig:sub_degree_mst}
	\end{subfigure}
	\begin{subfigure}{.45\textwidth}
		\centering
		\includegraphics[width=1\linewidth]{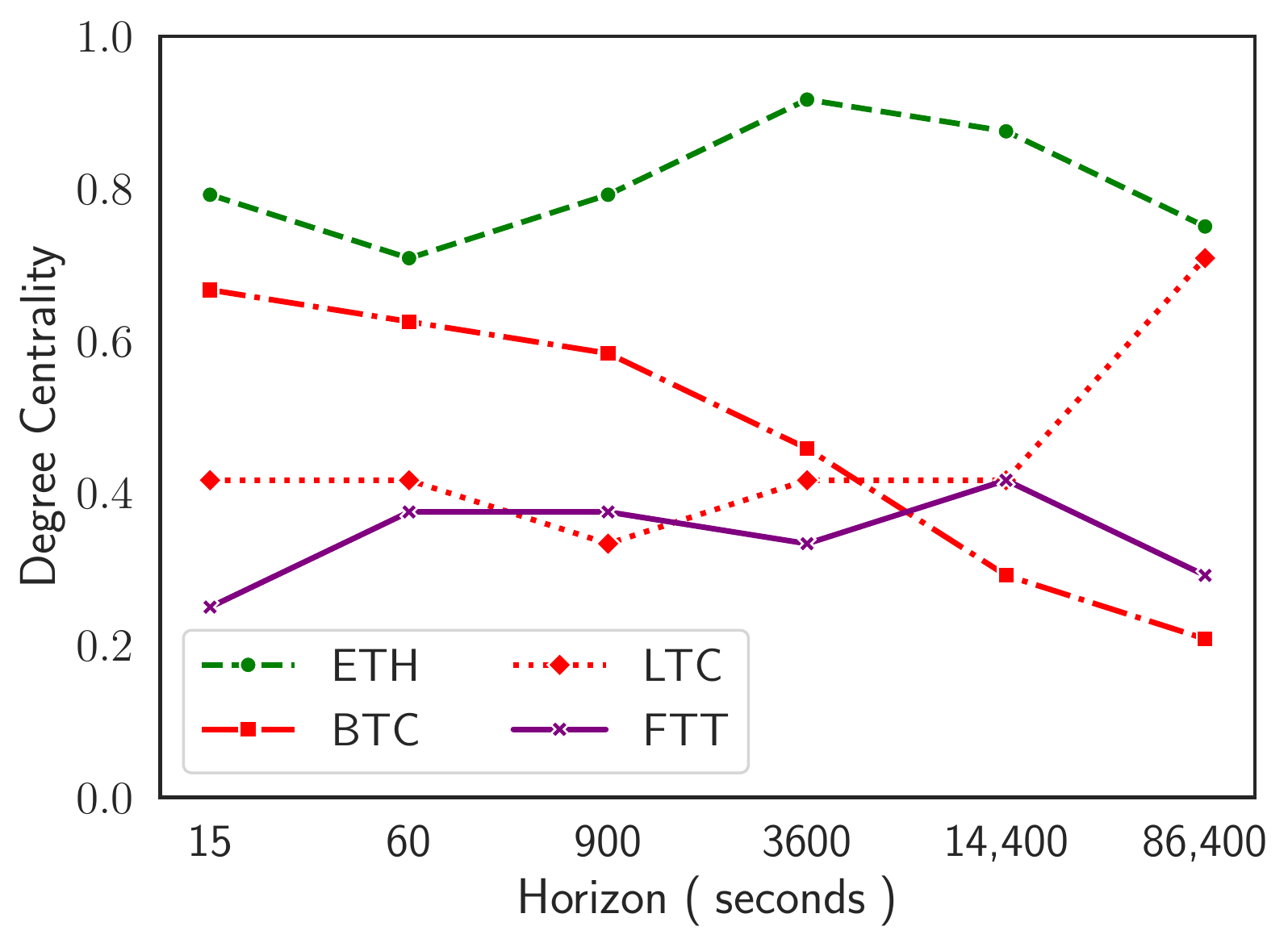}  
		\caption{\centering}
		\label{fig:sub_degree_pmfg}
	\end{subfigure}
	\vspace{2mm}
	\caption{Degree centrality %MDPI: Please use commas to separate thousands for numbers with five or more digits (not four digits) in the picture, e.g., “10000” should be “10,000”. CHANGE DONE.
 computed on MST (\textbf{a}) and on TMFG (\textbf{b}) as a function of time sampling resolution. Results on the TMFG highlight the switch in the reference roles of mainstream~cryptocurrencies.}
	\label{fig:Degree_Centrality}
\end{figure}

\begin{figure}[H]
	\begin{subfigure}{.45\textwidth}
		\centering
		\includegraphics[width=1\linewidth]{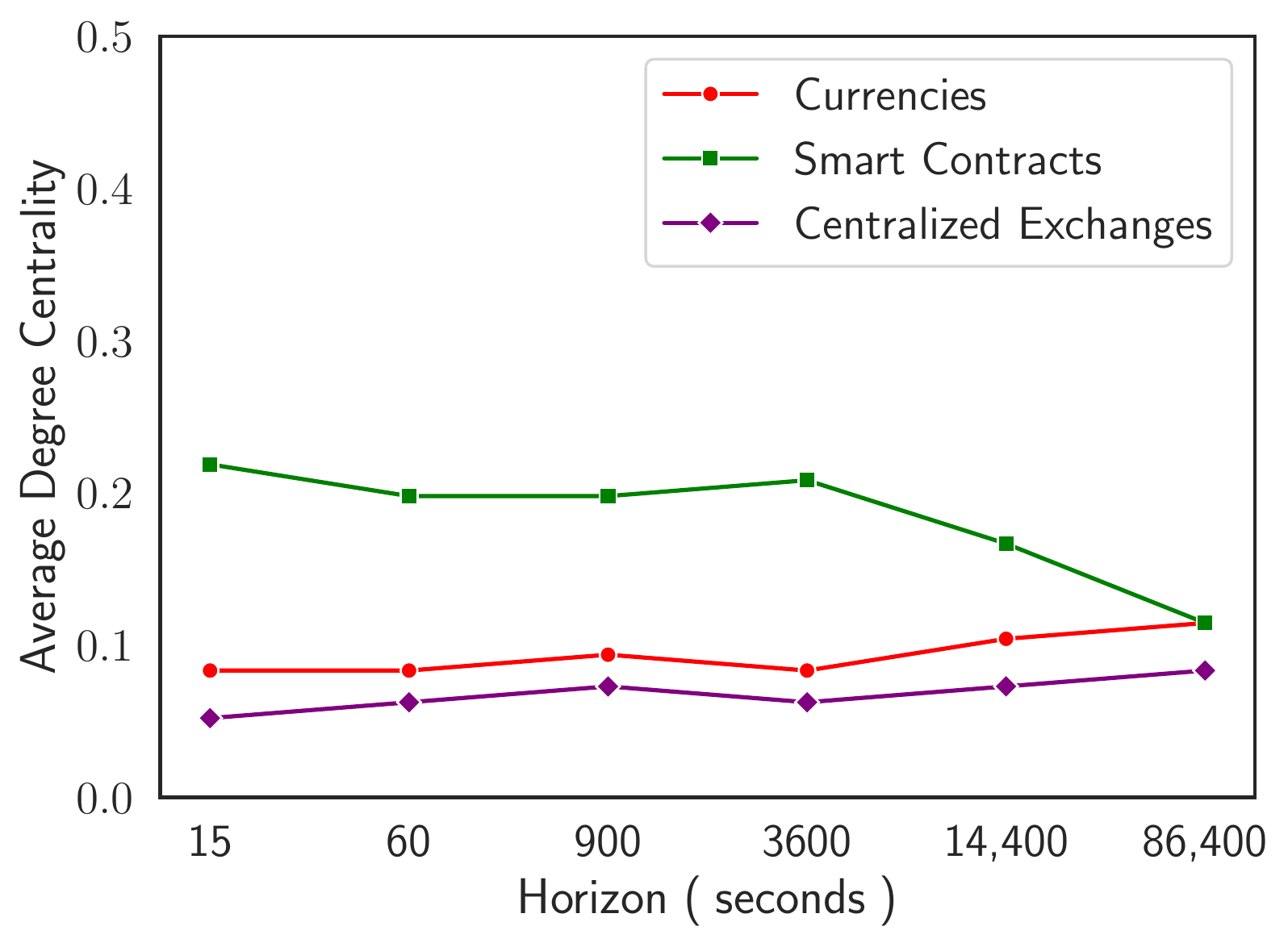}  
		\caption{\centering}
		\label{fig:sub_intra_sector_degree_mst}
	\end{subfigure}
	\begin{subfigure}{.45\textwidth}
		\centering
		\includegraphics[width=1\linewidth]{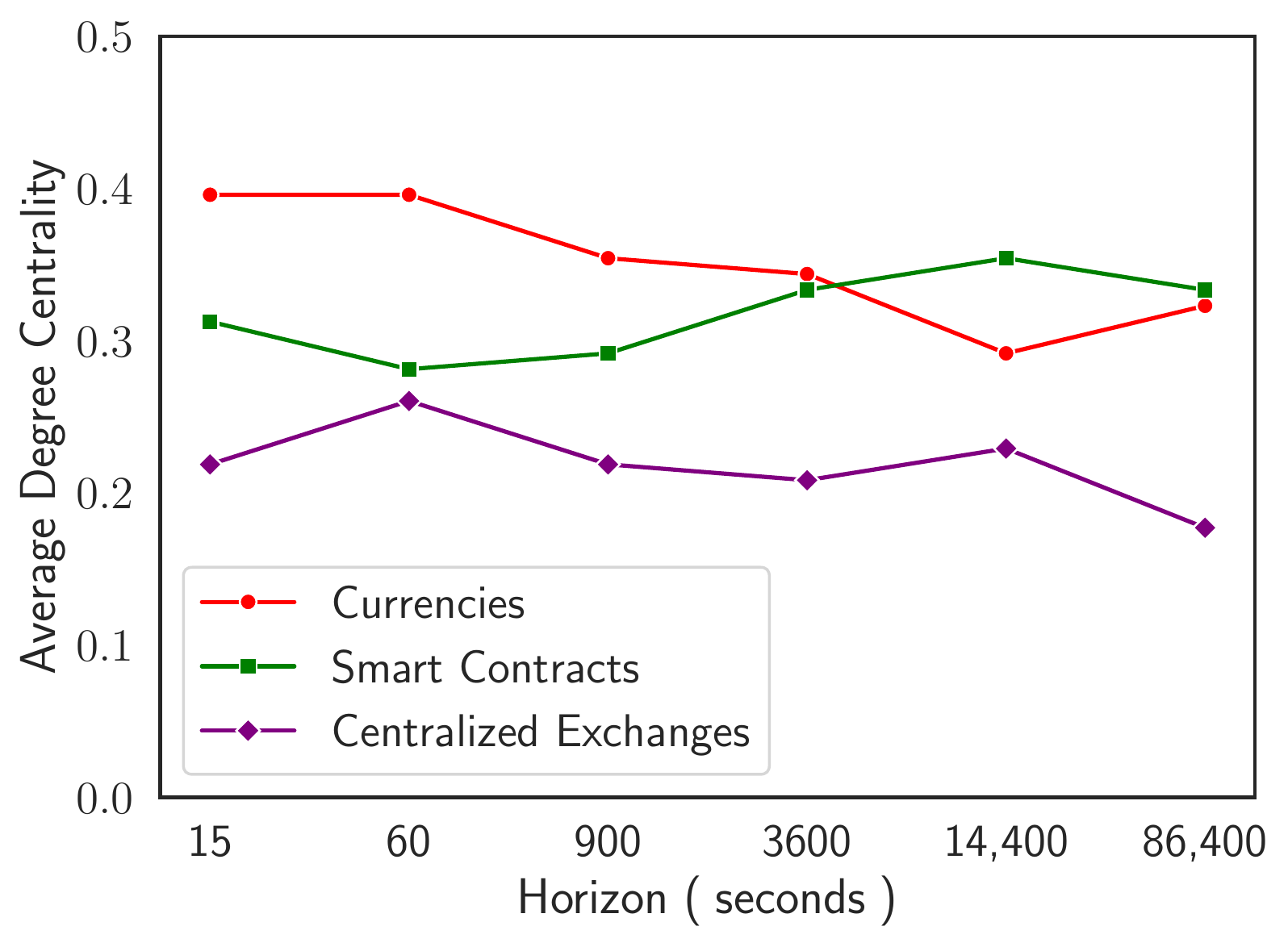}  
		\caption{\centering}
		\label{fig:sub_intra_sector_degree_tmfg}
	\end{subfigure}
	\vspace{2mm}
	\caption{Group degree centrality %MDPI: Please use commas to separate thousands for numbers with five or more digits (not four digits) in the picture, e.g., “10000” should be “10,000”. CHANGE DONE
 computed on MST (\textbf{a}) and on TMFG (\textbf{b}) for Currencies sector, Smart Contracts sector, and Centralized Exchanges sector. Group degree centrality of a set of nodes is defined as the fraction of non-group members connected to group members. Sectors are defined following the taxonomy by~\cite{Messari}.}
	\label{fig:Intra_Sector_Degree_Centrality}
\end{figure}

In the case of TMFG representations, it is harder to graphically detect similar dynamics. Figure~\ref{fig:Degree_Centrality} offers a comparative perspective between behaviours of the two network-based information filtering approaches. It shows horizon dependent evolutionary dynamics of degree centrality (i.e., measurement of the number of connections owned by a \linebreak node)~\cite{carrington2005models, kleinberg1999web, maharani2014degree} for Ethereum, Bitcoin, Litecoin, and FTX Token both for MST and for TMFG. Cross-assets similarities can be detected between the two types of graphs. In~the case of MSTs, degree centrality is less sensitive to minor changes in reference roles played by mainstream cryptocurrencies across time horizons, amplifying only `extreme' ones. In~the case of TMFGs, on~the contrary, the~same centrality measure is able to capture even small variations in network structure. This observation can be easily explained considering the amount of information the two representations are able to express. This study can be further extended looking at sectors of cryptocurrencies instead of at singular assets. Figure~\ref{fig:Intra_Sector_Degree_Centrality} reports the evolution of degree centrality for the three sectors Ethereum, Bitcoin, Litecoin, and FTX Token belong to: the Currencies sector, the~Smart Contracts sector, and the Centralized Exchanges sector. We remark that there is no consensus on a unique mapping between cryptocurrencies and sectors. The~taxonomy adopted in the current paper is described in~\cite{Messari} and corresponds to the one used by Kraken~\cite{Kraken}.

Figure~\ref{fig:Intra_Sector_Degree_Centrality}a shows how, in~the case of MST, the~average degree centrality for the Smart Contracts sector strongly decreases starting from time horizon $\Delta t = 3600$, following the trend of its leading representative: Ethereum cryptocurrency. The~Currencies sector, on~the other hand, does not experience a decreasing trend and tends to remain stable across time horizons with low level of oscillations. In~this case the loss of centrality of Bitcoin after time horizon $\Delta t = 900$, is immediately compensated by Litecoin, which reaches a hierarchical reference role at coarser time sampling resolutions. The~case of Centralized Exchanges sector is different. It is stable across time horizons, without~experiencing any change in intra-sector reference role dynamics and always following the behaviour of its main representative, FTX token (see Figure~\ref{fig:Degree_Centrality}a). This last finding can be explained considering the source of the data used in the current research work. As~explained in Section~\ref{sec:Data}, we fetch data from the FTX digital currency exchange. This can cause, on~the one hand an over-estimation of the role played by the exchange specific token, FTX Token, in~the whole ecology of the system under investigation, and, on~the other hand, can give a potentially biased stability to the sector the asset belongs~to.

\section{Conclusions}\label{sec:Conclusions}
We investigate how cryptocurrency market's dependency structures evolve passing from high to low frequency time sampling resolutions. Starting from the log-returns of 25~liquid cryptocurrencies traded on the FTX digital currency exchange at 6 different time horizons spanning from 15 s to 1 day, we investigate pairwise correlations demonstrating that cryptocurrency market has an ``Epps effect'' which is comparable to the one widely studied in the equity market. Indeed, we show that the average correlation among assets increases moving from high to low frequency time horizons and we demonstrate how this dynamic is even more evident grouping cryptocurrencies into sectors. Using the concept of power dissimilarity measure, we review the building process of two network-based information filtering approaches: MST and TMFG. If, on~the one hand, MST has been historically used in the description of dependency structures of different financial markets, on~the other hand, this is the very first time TMFG is used to study interactions between digital assets at different time scales. Studying topologies of MSTs at finer time sampling resolutions, we observe structurally simpler networks characterised by an hub-and-spoke configuration with statistically robust links. We observe an increase in the complexity of the networks' shape for coarser time sampling resolutions with a decrease in links' statistical robustness. Such an horizon-dependent structural change is reflected by the average path length of the networks, characterised by an increasing trend moving from high to low frequencies. TMFG offers a different perspective for the same problem. In~this case, we do not observe dramatic changes in networks' topologies across time horizons. Graphs are more compact and statistical robustness of links is maintained across time with negligible oscillations. As~a consequence of this, the~average path length is lower and almost constant across time horizons. Studying the relative position of assets in both MSTs and TMFGs through the usage of degree centrality measure, we outline the presence of multiple changes in the hierarchical reference role among the considered set of cryptocurrencies. These changes strongly characterise singular cryptocurrencies. We find that Ethereum acts as a hierarchical reference node for the majority of other assets and maintains this role across time, gradually losing its centrality at coarser time horizons. There is not a clear economic explanation for this result. We know that lots of other cryptocurrencies are based on the Ethereum's blockchain technology but we do not think this represents a sufficient explanation to our finding. Other cryptocurrencies play a similar role with respect to smaller clusters of assets at specific time horizons. We refer specifically to Bitcoin, Litecoin, and FTX Token. Differently from Ethereum, their role does not emerge at finer time sampling resolutions and should be considered as the result of a structured evolutionary process across time horizons. We conclude stating that sectors' dynamics captured by the chosen network-based information filtering approaches are poorly affected by the ones of their main representatives, efficiently absorbing horizon-dependent changes in cryptocurrencies dynamics. This is true especially for TMFG. Indeed, looking at the evolution of the degree centrality of the Smart Contracts and Currencies sectors, one can observe that dynamics captured by MST are strongly influenced by the ones of Ethereum and Bitcoin. This does not happen in the case of TMFG where sectors' dynamics are typically detached from the ones of specific~cryptocurrencies. 

%% optional
%\supplementary{The following supporting information can be downloaded at:  \linksupplementary{s1}, Figure S1: title; Table S1: title; Video S1: title.}

\vspace{6pt}

%%%%%%%%%%%%%%%%%%%%%%%%%%%%%%%%%%%%%%%%%%
\authorcontributions{Conceptualization, A.B. and T.A.; methodology, A.B. and T.A.; software, A.B.; validation, A.B. and T.A.; formal analysis, A.B. and T.A.; investigation, A.B. and T.A.; writing---original draft preparation, A.B.; writing---review and editing, T.A.; visualization, A.B.; supervision, T.A.; project administration, T.A.; funding acquisition, T.A. All authors have read and agreed to the published version of the~manuscript.}

\funding{This research was funded by ESRC (ES/K002309/1), EPSRC (EP/P031730/1) and EC (H2020-ICT-2018-2 825215).}
\institutionalreview{Not applicable} %MDPI: In this section, you should add the Institutional Review Board Statement and approval number, if relevant to your study. You might choose to exclude this statement if the study did not require ethical approval. Please note that the Editorial Office might ask you for further information. Please add “The study was conducted in accordance with the Declaration of Helsinki, and approved by the Institutional Review Board (or Ethics Committee) of NAME OF INSTITUTE (protocol code XXX and date of approval).” for studies involving humans. OR “The animal study protocol was approved by the Institutional Review Board (or Ethics Committee) of NAME OF INSTITUTE (protocol code XXX and date of approval).” for studies involving animals. OR “Ethical review and approval were waived for this study due to REASON (please provide a detailed justification).” OR “Not applicable” for studies not involving humans or animals.

\informedconsent{Not applicable} %MDPI: Any research article describing a study involving humans should contain this statement. Please add ``Informed consent was obtained from all subjects involved in the study.'' OR ``Patient consent was waived due to REASON (please provide a detailed justification).'' OR ``Not applicable'' for studies not involving humans. You might also choose to exclude this statement if the study did not involve humans.

%Written informed consent for publication must be obtained from participating patients who can be identified (including by the patients themselves). Please state ``Written informed consent has been obtained from the patient(s) to publish this paper'' if applicable.
\dataavailability{Data are accessible for free using the CCXT~\cite{CCXT} Python Package.} 

\acknowledgments{The authors acknowledge many members of the Financial Computing and Analytics Group at University College London. A~special thank to Silvia Bartolucci,  David Vidal-Tomás, and Yuanrong Wang. Additionally, thanks to Agne Kazakeviciute for fruitful discussions on foundational topics related to this~work.}

\conflictsofinterest{The authors declare no conflicts of interest. The~funders had no role in the design of the study; in the collection, analyses, or~interpretation of data; in the writing of the manuscript; or in the decision to publish the~results.} 

%%%%%%%%%%%%%%%%%%%%%%%%%%%%%%%%%%%%%%%%%%
%% Optional
%\sampleavailability{Samples of the compounds ... are available from the authors.}

%% Only for journal Encyclopedia
%\entrylink{The Link to this entry published on the encyclopedia platform.}

\abbreviations{Abbreviations}{
The following abbreviations are used in this manuscript:\\

\noindent 
\begin{tabular}{@{}ll}
DCE & Digital Currency Exchange\\
MST & Minimum Spanning Tree\\
PMFG & Planar Maximally Filtered Graph\\
TMFG & Triangulated Maximally Filtered Graph
\end{tabular}
}

%%%%%%%%%%%%%%%%%%%%%%%%%%%%%%%%%%%%%%%%%%
%% Optional
\appendixtitles{no} % Leave argument "no" if all appendix headings stay EMPTY (then no dot is printed after "Appendix A"). If~the appendix sections contain a heading then change the argument to "yes".
\appendixstart
\appendix
\section[\appendixname~\thesection]{}\label{Appendix_A}
\vspace{-10pt}

\begin{figure}[H]
	\hspace{-20pt}\begin{subfigure}{.5\textwidth}
		\centering
		\includegraphics[scale=0.3]{./images/MST_coloured_15}  
		\caption{\centering}
		\label{}
	\end{subfigure}
	\vspace{1mm}	
	\begin{subfigure}{.5\textwidth}
		\centering
		\includegraphics[scale=0.3]{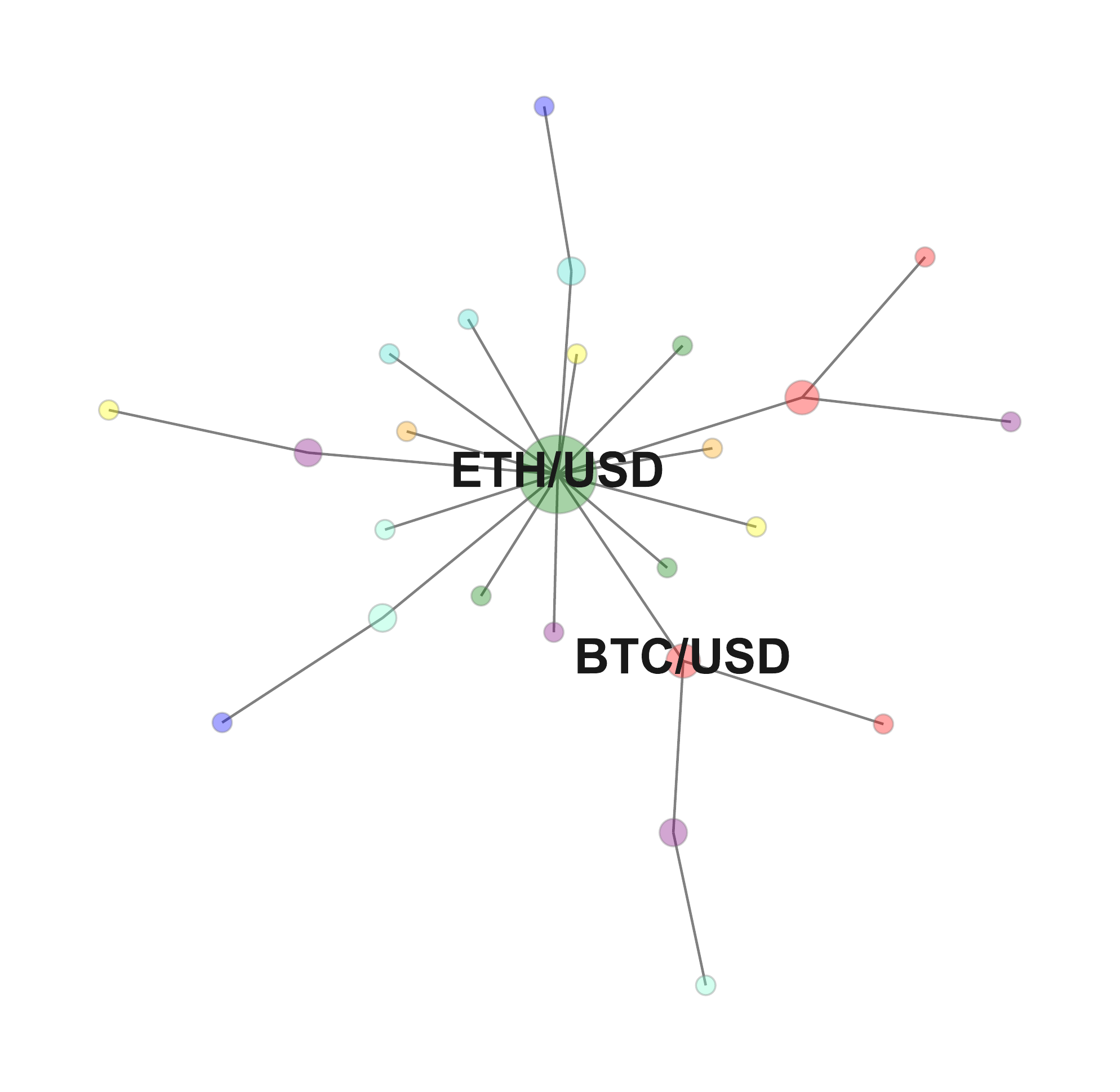}  
		\caption{\centering}
	\end{subfigure}
	\vspace*{1mm}
	\begin{subfigure}{.5\textwidth}
		\centering
		\includegraphics[scale=0.3]{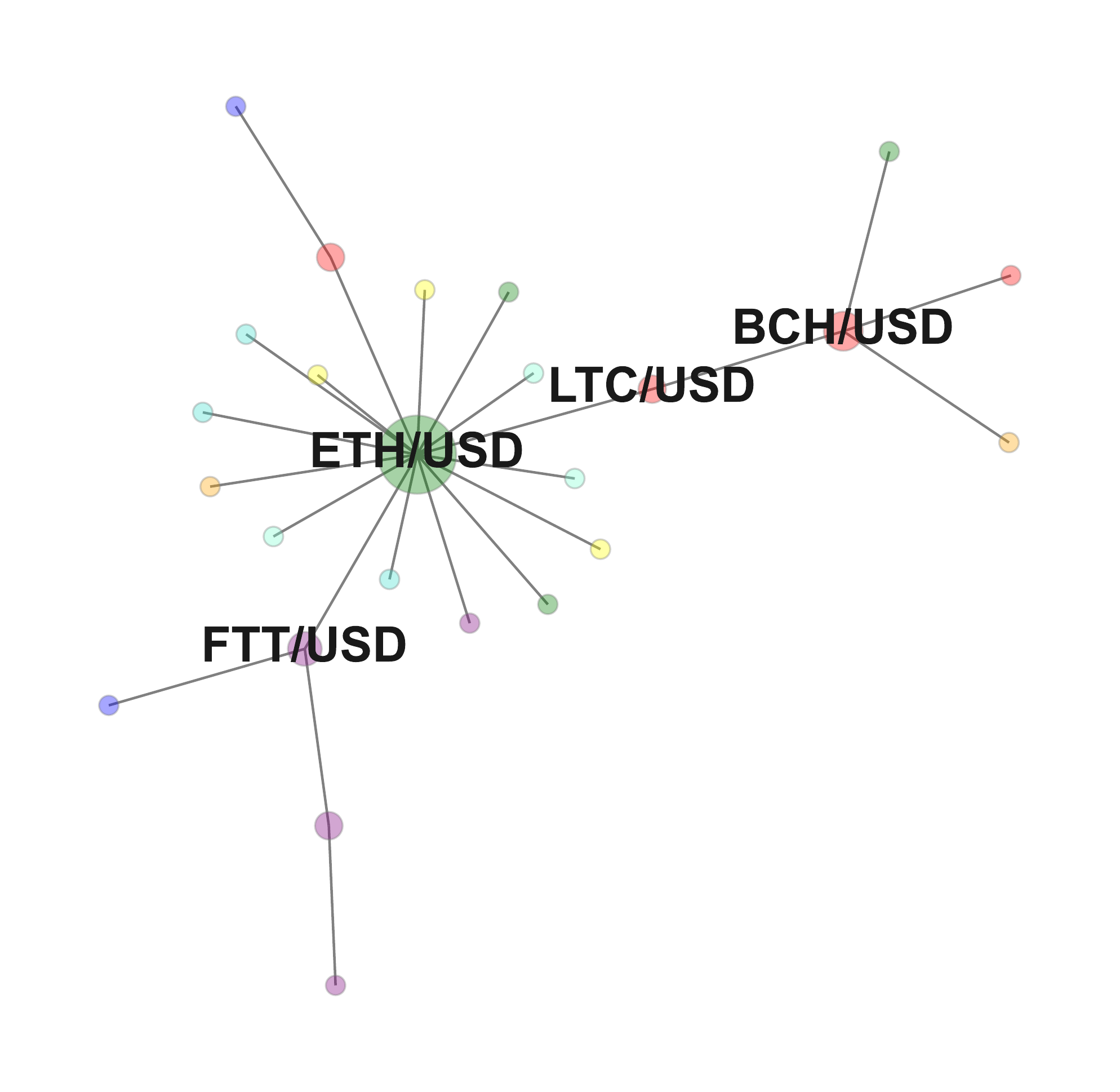}  
		\caption{\centering}
	\end{subfigure}
	\begin{subfigure}{.5\textwidth}
		\centering
		\includegraphics[scale=0.3]{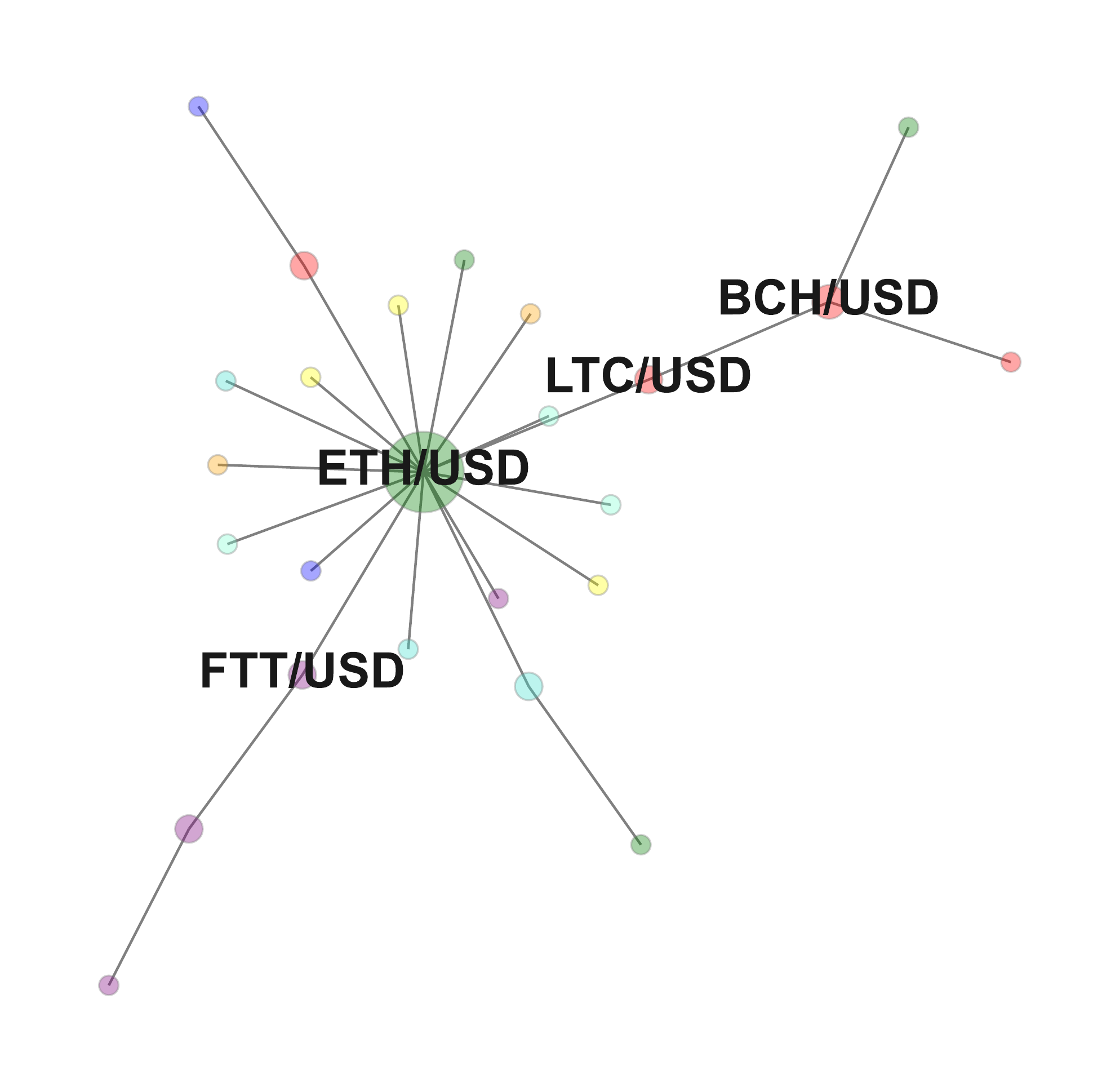}  
		\caption{\centering}
	\end{subfigure}
	\vspace{1mm}
	\begin{subfigure}{.5\textwidth}
		\centering
		\includegraphics[scale=0.3]{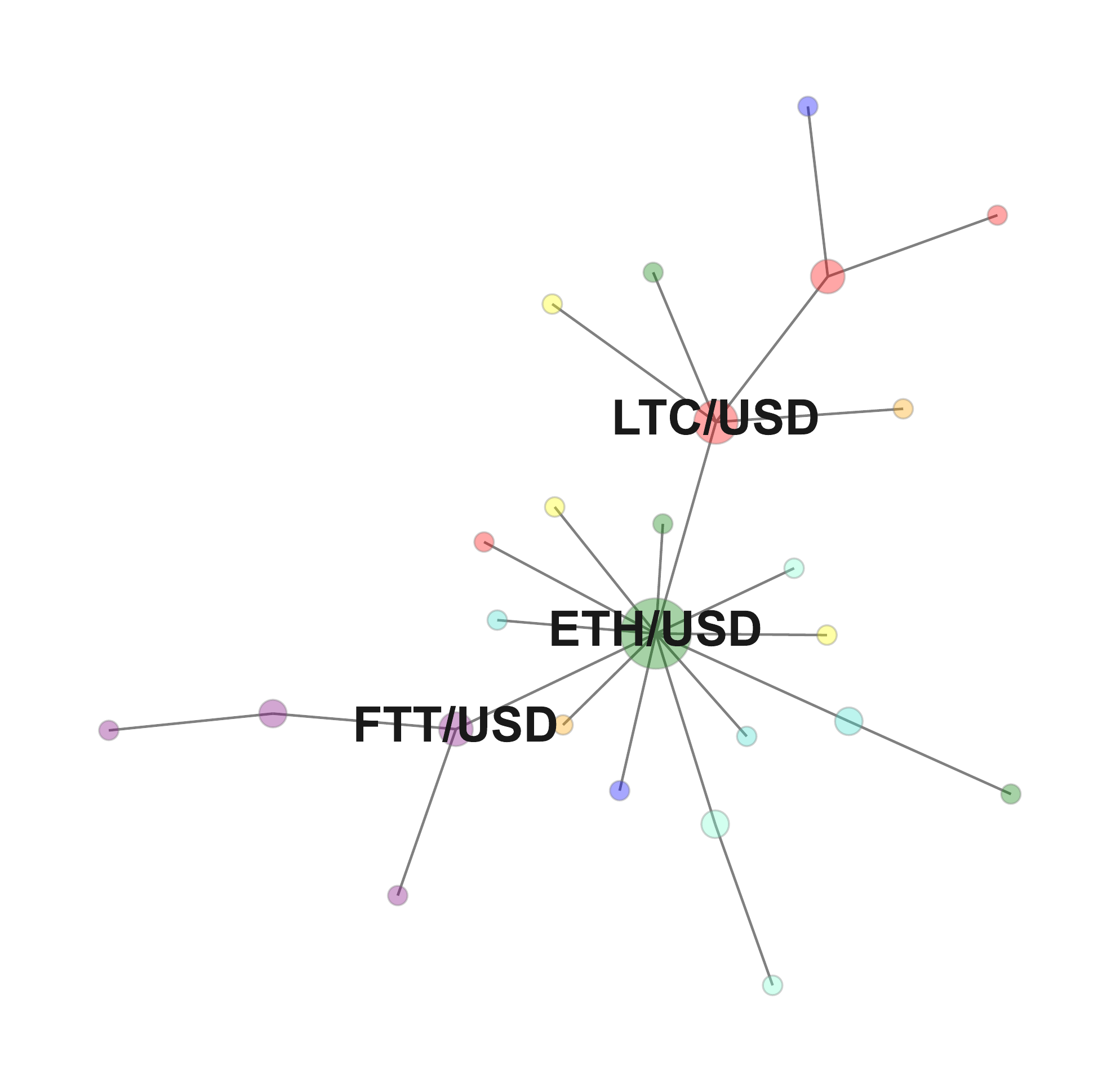}  
		\caption{\centering}
	\end{subfigure}
	\begin{subfigure}{.5\textwidth}
		\centering
		\includegraphics[scale=0.3]{./images/MST_coloured_86400}  
		\caption{\centering}
	\end{subfigure}
	
	\vspace*{1mm}
	
	\caption{Minimum Spanning Tree representing log-returns time series' dependency structure computed at (\textbf{a}) 15 s, (\textbf{b}) 1 min, (\textbf{c}) 15 min,  (\textbf{d}) 1 h,  (\textbf{e}) 4 h, and (\textbf{f}) 1 day. The~adopted colour mapping scheme follows the sectors' taxonomy by~\cite{Messari}: \textbf{red} $\rightarrow$ currencies, \textbf{green} $\rightarrow$ smart contract platforms, \textbf{blue} $\rightarrow$ stablecoins, \textbf{pink} $\rightarrow$ centralized exchanges, \textbf{orange} $\rightarrow$ scaling, \textbf{turquoise} $\rightarrow$ decentralized exchanges, \textbf{fuchsia}  $\rightarrow$ lending, and \textbf{yellow} $\rightarrow$ all the other sectors. Dashed, red edges represent negatives linear correlations among pairs of cryptocurrencies. Only{ hub} nodes are~labelled.}
\end{figure}

\section[\appendixname~\thesection]{}\label{Appendix_B}
\vspace{-10pt}
\begin{figure}[H]
	\hspace{-20pt}\begin{subfigure}{.5\textwidth}
		\centering
		\includegraphics[scale=0.3]{./images/TMFG_coloured_15}  
		\caption{\centering}
		\label{}
	\end{subfigure}
	\vspace*{1mm}
	\begin{subfigure}{.5\textwidth}
		\centering
		\includegraphics[scale=0.3]{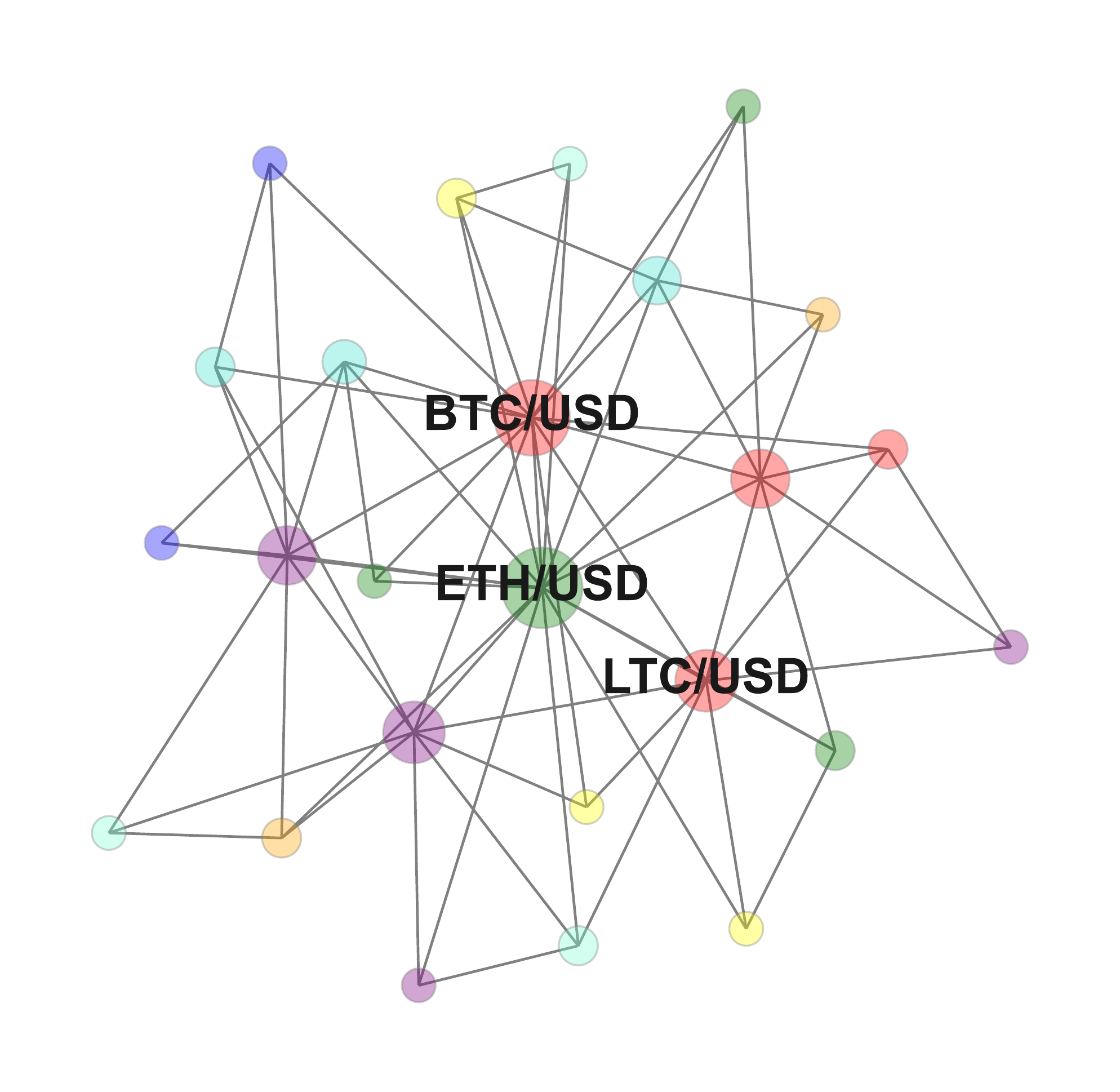}
		\caption{\centering}
	\end{subfigure}
	\vspace*{1mm}
	\begin{subfigure}{.5\textwidth}
		\centering
		\includegraphics[scale=0.3]{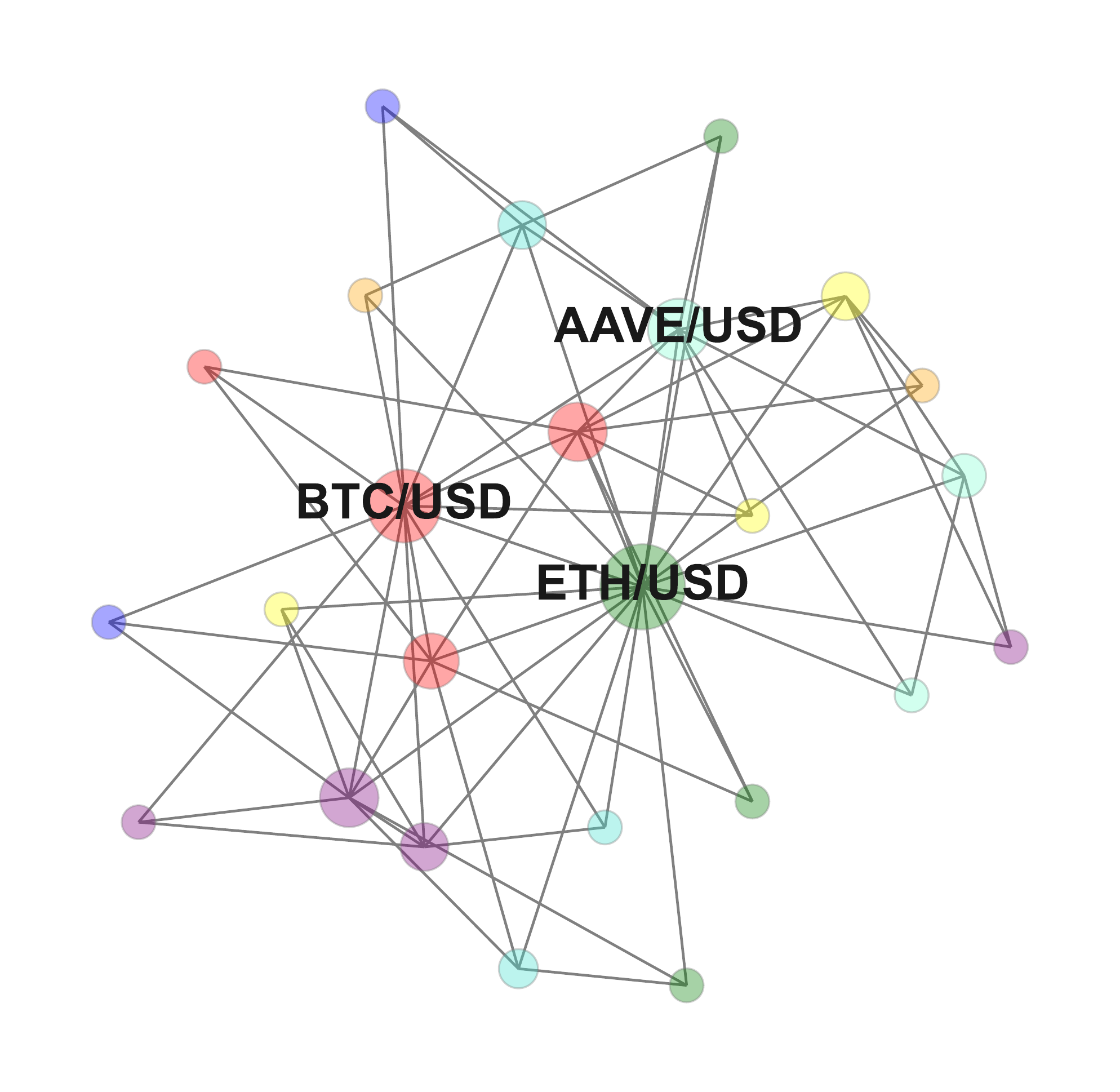}  
		\caption{\centering}
	\end{subfigure}
	\begin{subfigure}{.5\textwidth}
		\centering
		\includegraphics[scale=0.3]{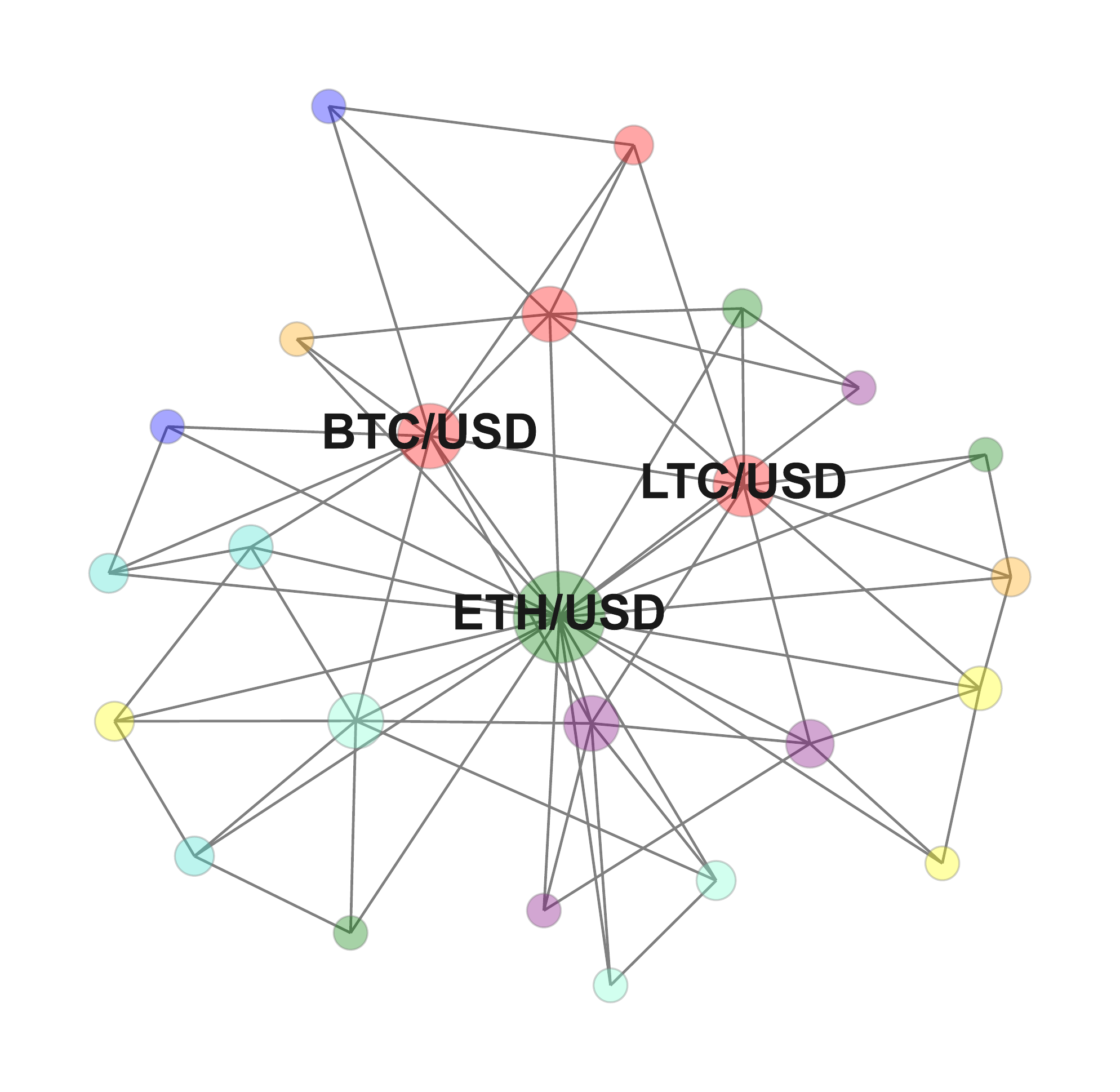}  
		\caption{\centering}
	\end{subfigure}
	\vspace*{1mm}
	\begin{subfigure}{.5\textwidth}
		\centering
		\includegraphics[scale=0.3]{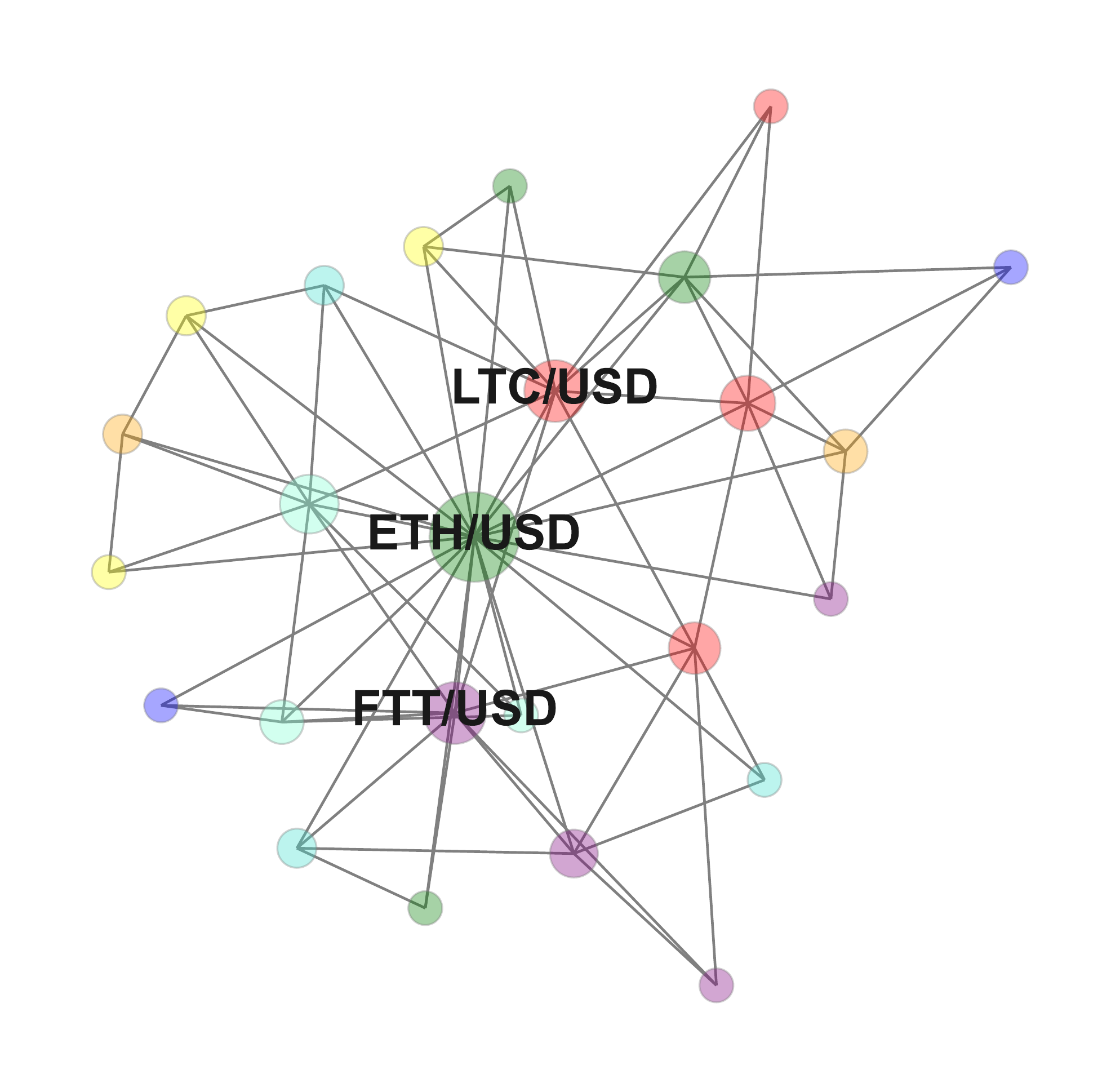}  
		\caption{\centering}
	\end{subfigure}
	\begin{subfigure}{.5\textwidth}
		\centering
		\includegraphics[scale=0.3]{./images/TMFG_coloured_86400}  
		\caption{\centering}
	\end{subfigure}
	
	\vspace{1mm}
	
	\caption{Triangulated Maximally Filtered Graphs representing log-returns time series' dependency structure computed at (\textbf{a}) 15 s, (\textbf{b}) 1 min, (\textbf{c}) 15 min,  (\textbf{d}) 1 h,  (\textbf{e}) 4 h, and  (\textbf{f}) 1 day. The~adopted colour mapping scheme follows the sectors' taxonomy by~\cite{Messari}: \textbf{red} $\rightarrow$ currencies, \textbf{green} $\rightarrow$ smart contract platforms, \textbf{blue} $\rightarrow$ stablecoins, \textbf{pink} $\rightarrow$ centralized exchanges, \textbf{orange} $\rightarrow$ scaling, \textbf{turquoise} $\rightarrow$ decentralized exchanges, \textbf{fuchsia}  $\rightarrow$ lending, and \textbf{yellow} $\rightarrow$ all the other sectors. Dashed, red edges represent negatives linear correlations among pairs of cryptocurrencies. Only {hub} nodes are~labelled.}
\end{figure}

\section[\appendixname~\thesection]{}\label{Appendix_C}

\begin{figure}[H]
	\begin{subfigure}{.5\textwidth}
		\centering
		\includegraphics[scale=0.17]{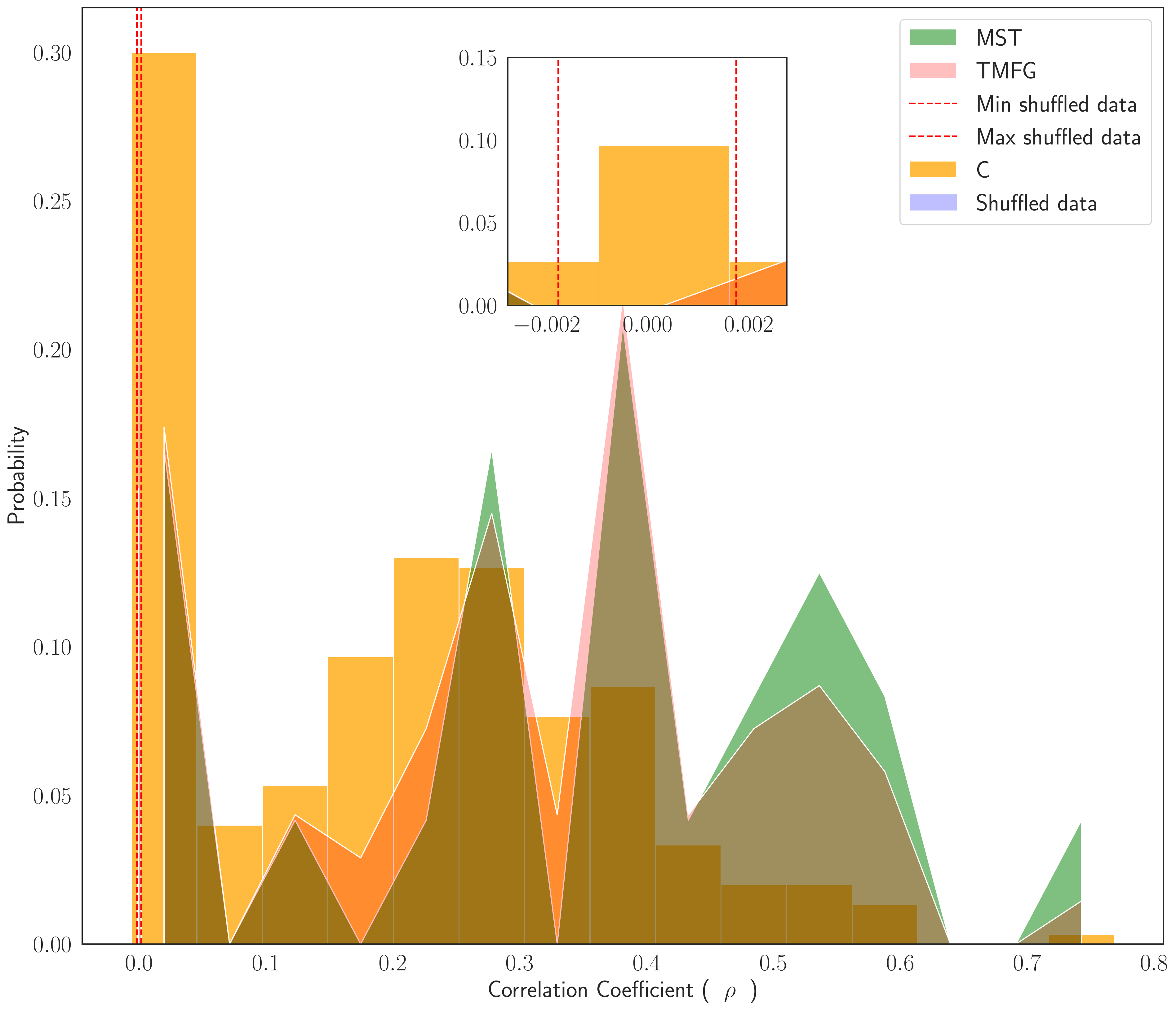}  
		\caption{\centering}
		\label{}
	\end{subfigure}
	\vspace*{1mm}
	\begin{subfigure}{.5\textwidth}
		\centering
		\includegraphics[scale=0.163]{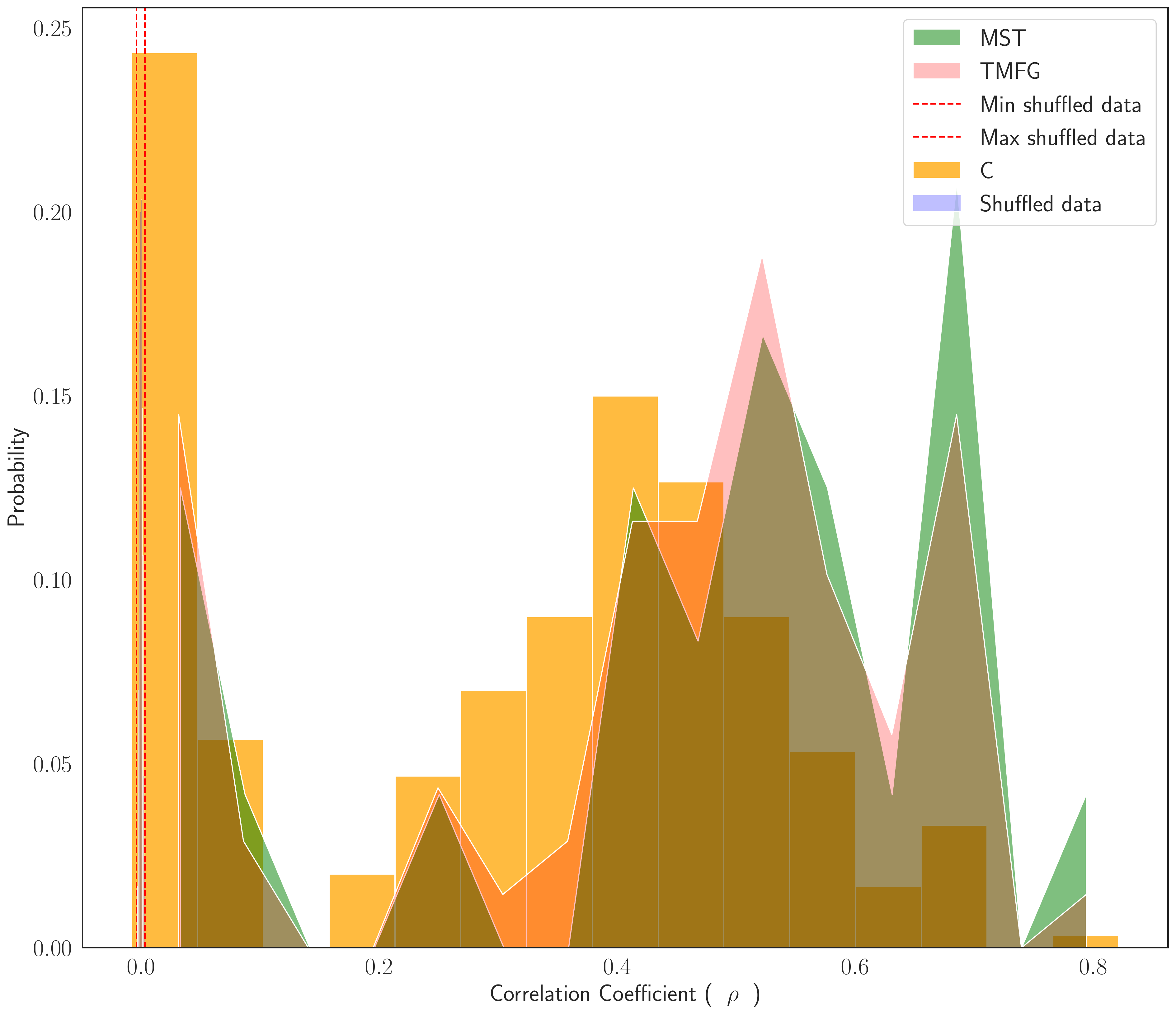}
		\caption{\centering}
	\end{subfigure}
	\vspace*{2mm}
	\begin{subfigure}{.5\textwidth}
		\centering
		\includegraphics[scale=0.163]{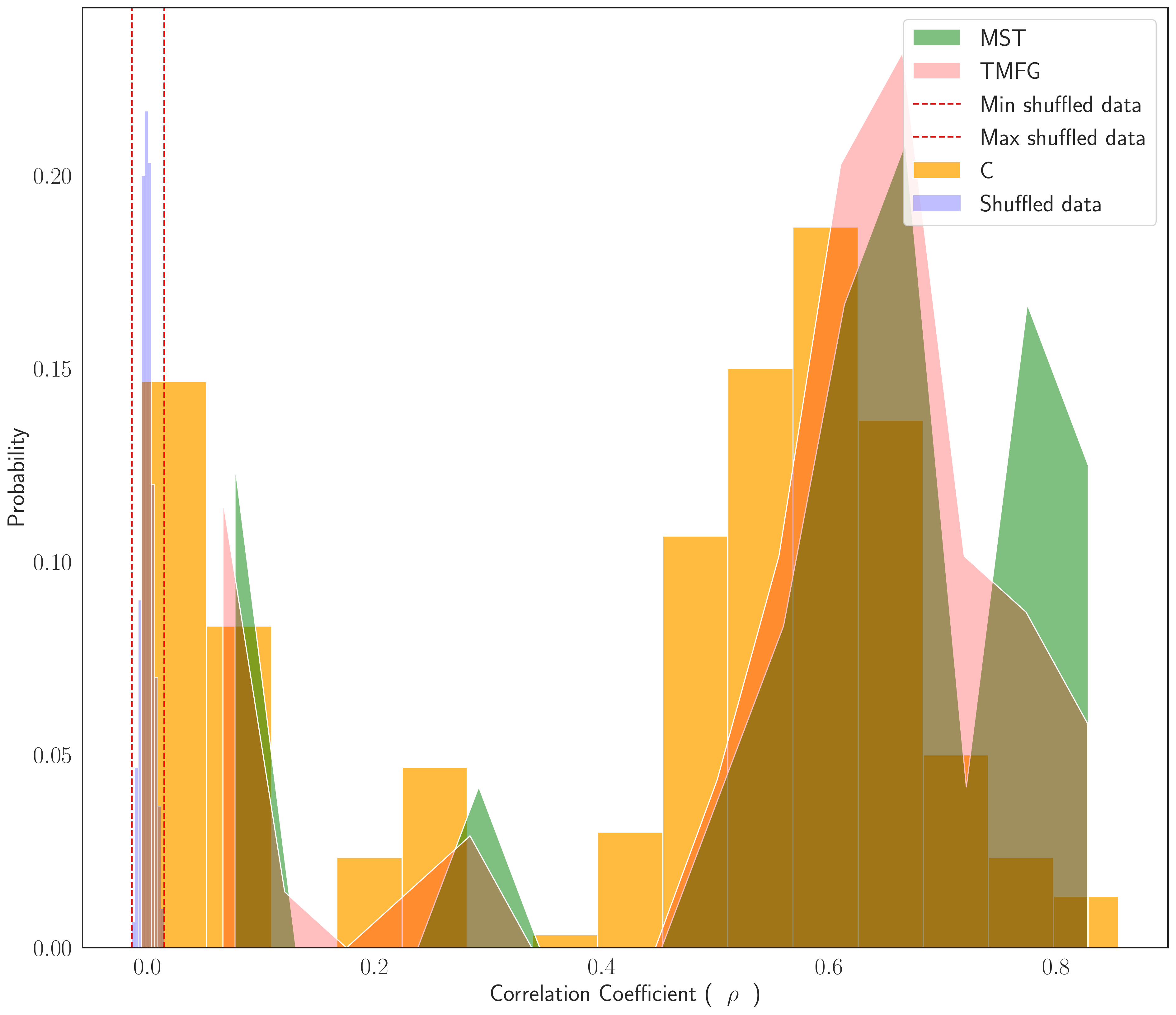}  
		\caption{\centering}
	\end{subfigure}
	\begin{subfigure}{.5\textwidth}
		\centering
		\includegraphics[scale=0.165]{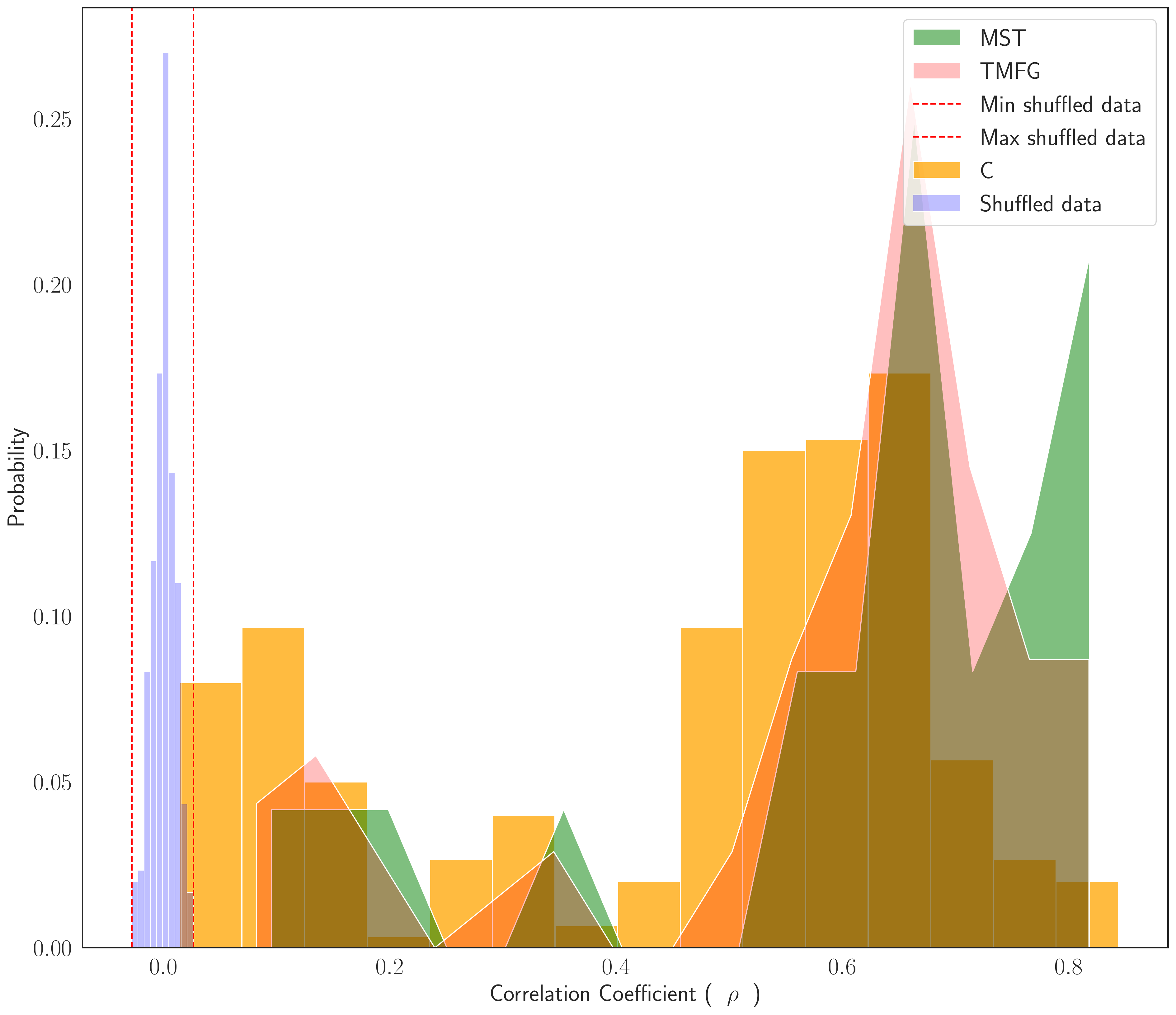}  
		\caption{\centering}
	\end{subfigure}
	\vspace*{2mm}
	\begin{subfigure}{.5\textwidth}
		\centering
		\includegraphics[scale=0.165]{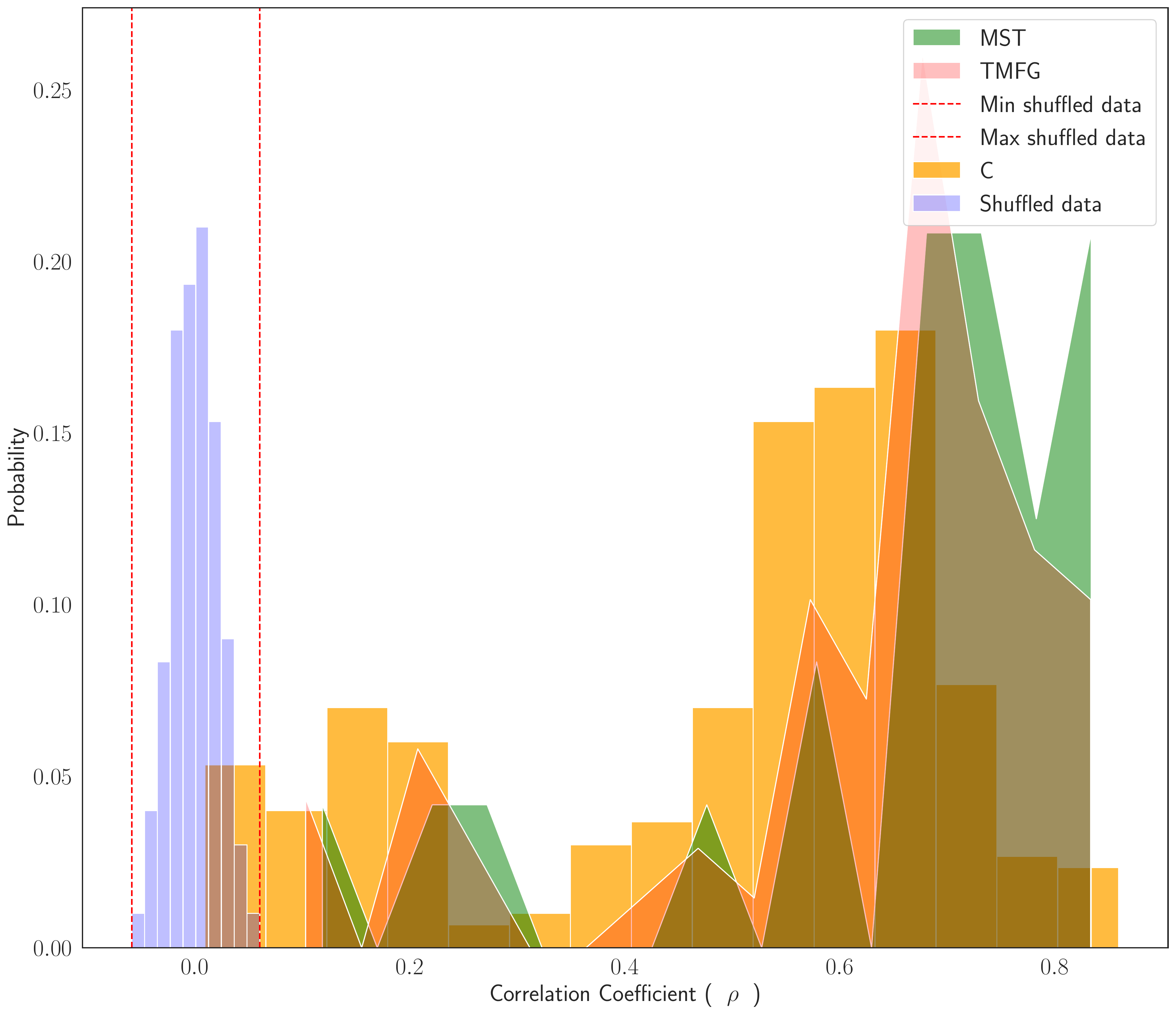}  
		\caption{\centering}
	\end{subfigure}
	\begin{subfigure}{.5\textwidth}
		\centering
		\includegraphics[scale=0.165]{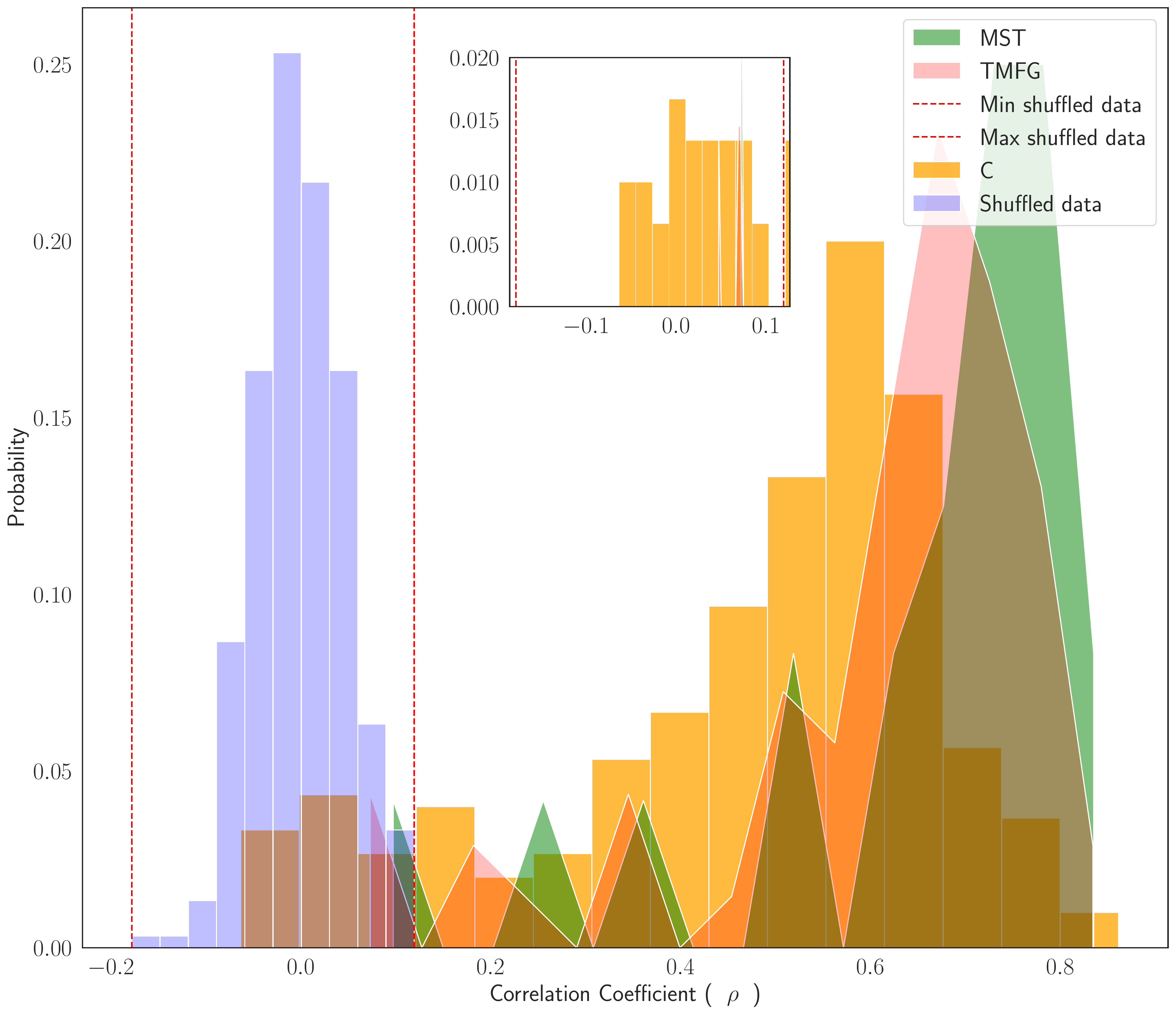}  
		\caption{\centering}
	\end{subfigure}
	
	\vspace{1mm}
	
	\caption{Probability distribution of correlation coefficients for the empirical correlation matrix \textbf{C}, MST, TMFG, and correlation matrix of shuffled log-returns time series computed at (\textbf{a}) 15 s, (\textbf{b}) 1 min, (\textbf{c}) 15 min, (\textbf{d}) 1 h, (\textbf{e}) 4 h, and (\textbf{f}) 1~day.}
\end{figure}

\section[\appendixname~\thesection]{}\label{Appendix_D}
\vspace{-6pt}
\begin{table}[H]
\tablesize{\small}
\caption{Number of links of the empirical correlation matrix \textbf{C}, of~the MST and of the TMFG having a value higher than the minimum and lower than the maximum correlation coefficient detected by shuffling log-returns time series at different time horizons. Shuffling operation is repeated 100 times. Results can be interpreted as p-values of the average correlation coefficient computed for \textbf{C}, for~the MST and for the~TMFG.}
\label{tab:Shuffling_Validation_Table}

\setlength{\cellWidtha}{\textwidth/4-2\tabcolsep-0in}
\setlength{\cellWidthb}{\textwidth/4-2\tabcolsep-0in}
\setlength{\cellWidthc}{\textwidth/4-2\tabcolsep-0in}
\setlength{\cellWidthd}{\textwidth/4-2\tabcolsep-0in}
\scalebox{1}[1]{\begin{tabularx}{\textwidth}{>{\centering\arraybackslash}m{\cellWidtha}>{\centering\arraybackslash}m{\cellWidthb}>{\centering\arraybackslash}m{\cellWidthc}>{\centering\arraybackslash}m{\cellWidthd}}
\toprule

\boldmath{$\Delta t$} & \textbf{C} & \textbf{MST} & \textbf{TMFG} \\ \midrule
15                  & 36         & 0            & 1             \\
60                  & 28         & 0            & 0             \\
900                 & 2          & 0            & 0             \\
3600                & 2          & 0            & 0             \\
14,400               & 16         & 0            & 0             \\
86,400               & 34         & 1            & 3             \\

\bottomrule
\end{tabularx}}

\end{table}

%\begin{table}[H] \label{tab:Shuffling_Validation_Table}
%\centering
%\caption{Number of links of the empirical correlation matrix \textbf{C}, of~the MST and of the TMFG having a value higher than the minimum and lower than the maximum correlation coefficient detected by shuffling log-returns time series at different time horizons. Shuffling operation is repeated 100 times. Results can be interpreted as p-values of the average correlation coefficient computed for \textbf{C}, for~the MST and for the~TMFG.}
%\begin{tabular}{@{}cccc@{}}
%\toprule
%\textbf{$\Delta t$} & \textbf{C} & \textbf{MST} & \textbf{TMFG} \\ \midrule
%15                  & 36         & 0            & 1             \\
%60                  & 28         & 0            & 0             \\
%900                 & 2          & 0            & 0             \\
%3600                & 2          & 0            & 0             \\
%14400               & 16         & 0            & 0             \\
%86400               & 34         & 1            & 3             \\ \bottomrule
%\end{tabular}
%\end{table}

%%%%%%%%%%%%%%%%%%%%%%%%%%%%%%%%%%%%%%%%%%
\begin{adjustwidth}{-\extralength}{0cm}

\reftitle{References}

\end{adjustwidth}

\begin{thebibliography}{999}

\bibitem[Anderson(2018)]{anderson2018economy}
Anderson, P.W.
\newblock {\em The Economy as an Evolving Complex System}; {CRC Press}:  Boca Raton, FL, USA, %newly added information, please confirm. CONFIRMED.
  2018.

\bibitem[Comerton-Forde and Putni{\c{n}}{\v{s}}(2015)]{comerton2015dark}
Comerton-Forde, C.; Putni{\c{n}}{\v{s}}, T.J.
\newblock Dark trading and price discovery.
\newblock {\em J. Financ. Econ.} {\bf 2015}, {\em 118},~70--92.

\bibitem[Briola \em{et~al.}(2021)Briola, Turiel, Marcaccioli, and
  Aste]{briola2021deep}
Briola, A.; Turiel, J.; Marcaccioli, R.; Aste, T.
\newblock Deep reinforcement learning for active high frequency trading.
\newblock {\em arXiv preprint} {\bf 2021}, arXiv:2101.07107.

\bibitem[Briola \em{et~al.}(2020)Briola, Turiel, and Aste]{briola2020deep}
Briola, A.; Turiel, J.; Aste, T.
\newblock Deep learning modeling of limit order book: A comparative
  perspective.
\newblock {\em arXiv preprint} {\bf 2020}, arXiv:2007.07319.

\bibitem[Lux and Marchesi(1999)]{lux1999scaling}
Lux, T.; Marchesi, M.
\newblock Scaling and criticality in a stochastic multi-agent model of a
  financial market.
\newblock {\em Nature} {\bf 1999}, {\em 397},~498--500.

\bibitem[Aste \em{et~al.}(2010)Aste, Shaw, and Di~Matteo]{aste2010correlation}
Aste, T.; Shaw, W.; Di~Matteo, T.
\newblock Correlation structure and dynamics in volatile markets.
\newblock {\em New J. Phys.} {\bf 2010}, {\em 12},~085009.

\bibitem[Goodell(2022)]{goodell2022tokens}
Goodell, G.
\newblock Tokens and Distributed Ledgers in Digital Payment Systems.
\newblock {\em arXiv preprint} {\bf 2022}, arXiv:2207.07530.

\bibitem[Fang \em{et~al.}(2022)Fang, Ventre, Basios, Kanthan, Martinez-Rego,
  Wu, and Li]{fang2022cryptocurrency}
Fang, F.; Ventre, C.; Basios, M.; Kanthan, L.; Martinez-Rego, D.; Wu, F.; Li,
  L.
\newblock Cryptocurrency trading: A comprehensive survey.
\newblock {\em Financ. Innov.} {\bf 2022}, {\em 8},~1--59.

\bibitem[Harwick(2016)]{harwick2016cryptocurrency}
Harwick, C.
\newblock Cryptocurrency and the problem of intermediation.  \emph{Indep. Rev.} \textbf{2016}, \emph{20}, 569--588.

\bibitem[Rose \em{et~al.}(2015)Rose et~al.]{rose2015evolution}
Rose, C.
\newblock The evolution of digital currencies: Bitcoin, a cryptocurrency
  causing a monetary revolution.
\newblock {\em Int. Bus. Econ. Res. J.}
  {\bf 2015}, {\em 14},~617--622.

\bibitem[Wasserman \em{et~al.}(1994)Wasserman, Faust,
  et~al.]{wasserman1994social}
Wasserman, S.; Faust, K.
\newblock \emph{Social Network Analysis: Methods and Applications}; Cambridge University Press: Cambridge, UK, {1994}. %CONFIRMED

\bibitem[Newman \em{et~al.}(2002)Newman, Watts, and Strogatz]{newman2002random}
Newman, M.E.; Watts, D.J.; Strogatz, S.H.
\newblock Random graph models of social networks.
\newblock {\em Proc. Natl. Acad. Sci. USA} {\bf 2002},
  {\em 99},~2566--2572.

\bibitem[Ronfeldt and Arquilla(2001)]{ronfeldt2001networks}
Ronfeldt, D.F.; Arquilla, J.
\newblock {\em Networks and Netwars}; {Rand}: Santa Monica, California, USA. %MDPI: please confirm the location. CONFIRMED
  2001.

\bibitem[Balcan \em{et~al.}(2009)Balcan, Hu, Goncalves, Bajardi, Poletto,
  Ramasco, Paolotti, Perra, Tizzoni, Van~den Broeck,
  et~al.]{balcan2009seasonal}
Balcan, D.; Hu, H.; Goncalves, B.; Bajardi, P.; Poletto, C.; Ramasco, J.J.;
  Paolotti, D.; Perra, N.; Tizzoni, M.; Van~den Broeck, W.;  et~al.
\newblock Seasonal transmission potential and activity peaks of the new
  influenza A (H1N1): A Monte Carlo likelihood analysis based on human
  mobility.
\newblock {\em BMC Med.} {\bf 2009}, {\em 7},~1--12.

\bibitem[Hufnagel \em{et~al.}(2004)Hufnagel, Brockmann, and
  Geisel]{hufnagel2004forecast}
Hufnagel, L.; Brockmann, D.; Geisel, T.
\newblock Forecast and control of epidemics in a globalized world.
\newblock {\em Proc. Natl. Acad. Sci. USA} {\bf 2004},
  {\em 101},~15124--15129.

\bibitem[Sporns \em{et~al.}(2005)Sporns, Tononi, and
  K{\"o}tter]{sporns2005human}
Sporns, O.; Tononi, G.; K{\"o}tter, R.
\newblock The human connectome: A structural description of the human brain.
\newblock {\em PLoS Comput. Biol.} {\bf 2005}, {\em 1},~e42.

\bibitem[Hopkins(2007)]{hopkins2007network}
Hopkins, A.L.
\newblock Network pharmacology.
\newblock {\em Nat. Biotechnol.} {\bf 2007}, {\em 25},~1110--1111.

\bibitem[Wu \em{et~al.}(2008)Wu, Waber, Aral, Brynjolfsson, and
  Pentland]{wu2008mining}
Wu, L.; Waber, B.N.; Aral, S.; Brynjolfsson, E.; Pentland, A.
\newblock \emph{Mining Face-to-Face Interaction Networks Using Sociometric Badges:
  Predicting Productivity in an It Configuration Task}; {\bf2008}; %MDPI: please add the publisher and location of it. INFORMATION NOT FOUND.
\newblock {SSRN 1130251}.

\bibitem[Mantegna(1999)]{mantegna1999hierarchical}
Mantegna, R.N.
\newblock Hierarchical structure in financial markets.
\newblock {\em  Eur. Phys. J. B Condens. Matter Complex Syst.} {\bf 1999}, {\em 11},~193--197.

\bibitem[Bonanno \em{et~al.}(2003)Bonanno, Caldarelli, Lillo, and
  Mantegna]{bonanno2003topology}
Bonanno, G.; Caldarelli, G.; Lillo, F.; Mantegna, R.N.
\newblock Topology of correlation-based minimal spanning trees in real and
  model markets.
\newblock {\em Phys. Rev. E} {\bf 2003}, {\em 68},~046130.

\bibitem[Bonanno \em{et~al.}(2004)Bonanno, Caldarelli, Lillo, Micciche,
  Vandewalle, and Mantegna]{bonanno2004networks}
Bonanno, G.; Caldarelli, G.; Lillo, F.; Micciche, S.; Vandewalle, N.; Mantegna,
  R.N.
\newblock Networks of equities in financial markets.
\newblock {\em  Eur. Phys. J. B} {\bf 2004}, {\em 38},~363--371.

\bibitem[Bonanno \em{et~al.}(2001)Bonanno, Lillo, and
  Mantegna]{bonanno2001high}
Bonanno, G.; Lillo, F.; Mantegna, R.N.
\newblock High-frequency cross-correlation in a set of stocks. \emph{Quantitative Finance}  {\bf 2001}, 1:1, 96--104. %ADJUSTED

\bibitem[Wang and Aste(2021)]{wang2021dynamic}
Wang, Y.; Aste, T.
\newblock Dynamic Portfolio Optimization with Inverse Covariance Clustering.
\newblock {\em arXiv preprint} {\bf 2021}, arXiv:2112.15499.

\bibitem[Procacci and Aste(2021)]{procacci2021portfolio}
Procacci, P.F.; Aste, T.
\newblock Portfolio Optimization with Sparse Multivariate Modelling.
\newblock {\em arXiv preprint} {\bf 2021}, arXiv:2103.15232.

\bibitem[Wang and Aste(2022)]{wang2022sparsification}
Wang, Y.; Aste, T.
\newblock Sparsification and Filtering for Spatial-temporal GNN in Multivariate
  Time-series.
\newblock {\em arXiv preprint} {\bf 2022}, arXiv:2203.03991.

\bibitem[West \em{et~al.}(2001)West et~al.]{west2001introduction}
West, D.B.
\newblock {\em Introduction to Graph Theory}; Prentice Hall: Upper
  Saddle River, NJ,  %MDPI: please add the location of it. CHANGED
  2001; Volume~2. 

\bibitem[Massara \em{et~al.}(2017)Massara, Di~Matteo, and
  Aste]{massara2017network}
Massara, G.P.; Di~Matteo, T.; Aste, T.
\newblock Network filtering for big data: Triangulated maximally filtered
  graph.
\newblock {\em J. Complex Netw.} {\bf 2017}, {\em 5},~161--178.

\bibitem[Kra()]{Kraken}
Kraken: Bitcoin \& Cryptocurrency Exchange. Available online:  
\newblock \url{https://www.kraken.com}
\newblock (accessed on 22 March 2022).

\bibitem[Bonanno \em{et~al.}(2000)Bonanno, Vandewalle, and
  Mantegna]{bonanno2000taxonomy}
Bonanno, G.; Vandewalle, N.; Mantegna, R.N.
\newblock Taxonomy of stock market indices.
\newblock {\em Phys. Rev. E} {\bf 2000}, {\em 62},~R7615.

\bibitem[Epps(1979)]{epps1979comovements}
Epps, T.W.
\newblock Comovements in stock prices in the very short run.
\newblock {\em J. Am. Stat. Assoc.} {\bf 1979},
  {\em 74},~291--298.

\bibitem[Laloux \em{et~al.}(1999)Laloux, Cizeau, Bouchaud, and
  Potters]{laloux1999noise}
Laloux, L.; Cizeau, P.; Bouchaud, J.P.; Potters, M.
\newblock Noise dressing of financial correlation matrices.
\newblock {\em Phys. Rev. Lett.} {\bf 1999}, {\em 83},~1467.

\bibitem[Plerou \em{et~al.}(2002)Plerou, Gopikrishnan, Rosenow, Amaral, Guhr,
  and Stanley]{plerou2002random}
Plerou, V.; Gopikrishnan, P.; Rosenow, B.; Amaral, L.A.N.; Guhr, T.; Stanley,
  H.E.
\newblock Random matrix approach to cross correlations in financial data.
\newblock {\em Phys. Rev. E} {\bf 2002}, {\em 65},~066126.

\bibitem[Campbell \em{et~al.}(1997)Campbell, Lo, and
  MacKinlay]{campbell1997econometrics}
Campbell, J.Y.; Lo, A.; MacKinlay, C.
\newblock \emph{The Econometrics of Financial Markets}; Princeton University Press:
  Princeton, NJ, USA, {1997}.

\bibitem[Aste \em{et~al.}(2005)Aste, Di~Matteo, and Hyde]{aste2005complex}
Aste, T.; Di~Matteo, T.; Hyde, S.
\newblock Complex networks on hyperbolic surfaces.
\newblock {\em Phys. A Stat. Mech. Appl.} {\bf
  2005}, {\em 346},~20--26.

\bibitem[Vidal-Tom{\'a}s(2021)]{vidal2021entry}
Vidal-Tom{\'a}s, D.
\newblock The entry and exit dynamics of the cryptocurrency market.
\newblock {\em Res. Int. Bus. Financ.} {\bf 2021}, {\em
  58},~101504.

\bibitem[Kenett \em{et~al.}(2011)Kenett, Shapira, Madi, Bransburg-Zabary,
  Gur-Gershgoren, and Ben-Jacob]{kenett2011index}
Kenett, D.Y.; Shapira, Y.; Madi, A.; Bransburg-Zabary, S.; Gur-Gershgoren, G.;
  Ben-Jacob, E.
\newblock Index cohesive force analysis reveals that the US market became prone
  to systemic collapses since 2002.
\newblock {\em PLoS ONE} {\bf 2011}, {\em 6},~e19378.

\bibitem[Zikeba \em{et~al.}(2019)Zikeba, Kokoszczynski, and
  Sledziewska]{zikeba2019shock}
Zikeba, D.; Kokoszczynski, R.; Sledziewska, K.
\newblock Shock transmission in the cryptocurrency market. Is Bitcoin the most
  influential?
\newblock {\em Int. Rev. Financ. Anal.} {\bf 2019}, {\em
  64},~102--125.

\bibitem[Katsiampa \em{et~al.}(2021)Katsiampa, Yarovaya, and
  Zikeba]{katsiampa2021high}
Katsiampa, P.; Yarovaya, L.; Zikeba, D.
\newblock \emph{High-Frequency Connectedness between Bitcoin and Other Top-Traded
  Crypto Assets during the COVID-19 Crisis}; Elsevier: Amsterdam, The Netherlands. {2021}; %MDPI: please confirm the publisher and location. CONFIRMED
  SSRN 3871405.


\bibitem[Vidal-Tom{\'a}s(2022)]{vidal2022all}
Vidal-Tom{\'a}s, D.
\newblock All the frequencies matter in the Bitcoin market: An efficiency
  analysis.
\newblock {\em Appl. Econ. Lett.} {\bf 2022}, {\em 29},~212--218.

\bibitem[FTX()]{FTX}
FTX: FTX Cryptocurrency Derivatives Exchange. Available online:  
\newblock \url{https://ftx.com}.
\newblock (accessed on 22 March 2022).

\bibitem[CCX()]{CCXT}
Ccxt: CCXT---CryptoCurrency eXchange Trading Library. Available online:  
\newblock \url{https://github.com/ccxt/ccxt}
\newblock (accessed on 22 March 2022).

\bibitem[Mes()]{Messari}
Messari: Messari Crypto Research, Data, and Tools. Available online:  
\newblock \url{https://messari.io}
\newblock (accessed on 22 March 2022).

\bibitem[Dickey and Fuller(1979)]{dickey1979distribution}
Dickey, D.A.; Fuller, W.A.
\newblock Distribution of the estimators for autoregressive time series with a
  unit root.
\newblock {\em J. Am. Stat. Assoc.} {\bf 1979},
  {\em 74},~427--431.

\bibitem[Gower(1966)]{gower1966some}
Gower, J.C.
\newblock Some distance properties of latent root and vector methods used in
  multivariate analysis.
\newblock {\em Biometrika} {\bf 1966}, {\em 53},~325--338.

\bibitem[Soramaki \em{et~al.}(2016)Soramaki, Cook, and
  Laubsch]{soramaki2016network}
Soramaki, K.; Cook, S.; Laubsch, A.
\newblock A network-based method for visual identification of systemic risks.
\newblock {\em J. Netw. Theory Financ.} {\bf 2016}, {\em
  2},~67--101.

\bibitem[Mantegna and Stanley(1999)]{mantegna1999introduction}
Mantegna, R.N.; Stanley, H.E.
\newblock {\em Introduction to Econophysics: Correlations and Complexity in
  Finance}; Cambridge University Press:  Cambridge, UK, 1999. %CONFIRMED

\bibitem[Prim(1957)]{prim1957shortest}
Prim, R.C.
\newblock Shortest connection networks and some generalizations.
\newblock {\em  Bell Syst. Tech. J.} {\bf 1957}, {\em
  36},~1389--1401.

\bibitem[Tumminello \em{et~al.}(2005)Tumminello, Aste, Di~Matteo, and
  Mantegna]{tumminello2005tool}
Tumminello, M.; Aste, T.; Di~Matteo, T.; Mantegna, R.N.
\newblock A tool for filtering information in complex systems.
\newblock {\em Proc. Natl. Acad. Sci. USA} {\bf 2005},
  {\em 102},~10421--10426.

\bibitem[Aste and Di~Matteo(2006)]{aste2006dynamical}
Aste, T.; Di~Matteo, T.
\newblock Dynamical networks from correlations.
\newblock {\em Phys. A Stat. Mech. Appl.} {\bf
  2006}, {\em 370},~156--161.

\bibitem[Tumminello \em{et~al.}(2007)Tumminello, Di~Matteo, Aste, and
  Mantegna]{tumminello2007correlation}
Tumminello, M.; Di~Matteo, T.; Aste, T.; Mantegna, R.N.
\newblock Correlation based networks of equity returns sampled at different
  time horizons.
\newblock {\em  Eur. Phys. J. B} {\bf 2007}, {\em 55},~209--217.

\bibitem[Di~Matteo \em{et~al.}(2010)Di~Matteo, Pozzi, and Aste]{di2010use}
Di~Matteo, T.; Pozzi, F.; Aste, T.
\newblock The use of dynamical networks to detect the hierarchical organization
  of financial market sectors.
\newblock {\em  Eur. Phys. J. B} {\bf 2010}, {\em 73},~3--11.

\bibitem[Turiel \em{et~al.}(2020)Turiel, Barucca, and
  Aste]{turiel2020simplicial}
Turiel, J.D.; Barucca, P.; Aste, T.
\newblock Simplicial persistence of financial markets: Filtering, generative
  processes and portfolio risk.
\newblock {\em arXiv preprint} {\bf 2020}, arXiv:2009.08794.

\bibitem[Barfuss \em{et~al.}(2016)Barfuss, Massara, Di~Matteo, and
  Aste]{barfuss2016parsimonious}
Barfuss, W.; Massara, G.P.; Di~Matteo, T.; Aste, T.
\newblock Parsimonious modeling with information filtering networks.
\newblock {\em Phys. Rev. E} {\bf 2016}, {\em 94},~062306.

\bibitem[Tumminello \em{et~al.}(2007)Tumminello, Coronnello, Lillo, Micciche,
  and Mantegna]{tumminello2007spanning}
Tumminello, M.; Coronnello, C.; Lillo, F.; Micciche, S.; Mantegna, R.N.
\newblock Spanning trees and bootstrap reliability estimation in
  correlation-based networks.
\newblock {\em Int. J. Bifurc. Chaos} {\bf 2007}, {\em
  17},~2319--2329.

\bibitem[Carrington \em{et~al.}(2005)Carrington, Scott, and
  Wasserman]{carrington2005models}
Carrington, P.J.; Scott, J.; Wasserman, S.
\newblock {\em Models and Methods in Social Network Analysis}; 
  Cambridge University Press:  Cambridge, UK,  2005; Volume~28. %CONFIRMED

\bibitem[Kleinberg \em{et~al.}(1999)Kleinberg, Kumar, Raghavan, Rajagopalan,
  and Tomkins]{kleinberg1999web}
Kleinberg, J.M.; Kumar, R.; Raghavan, P.; Rajagopalan, S.; Tomkins, A.S.
\newblock The web as a graph: Measurements, models, and methods.
\newblock In \emph{International Computing and Combinatorics
  Conference}; Springer:  Berlin/Heidelberg, Germany,  1999; pp. 1--17. %CONFIRMED

\bibitem[Maharani \em{et~al.}(2014)Maharani, Gozali,
  et~al.]{maharani2014degree}
Maharani, W.; Gozali, A.A.
\newblock Degree centrality and eigenvector centrality in twitter.
\newblock In Proceedings of the 2014 8th International Conference on
  Telecommunication Systems Services and Applications (TSSA),  Kuta, Indonesia, 23--24 October 2014; pp.
  1--5. %CONFIRMED

\end{thebibliography}
\end{document}